\documentclass[manuscript,natbib=false]{acmart}

\usepackage{amsmath,amsfonts}
\usepackage{algorithm}
\usepackage{array}
\usepackage[caption=false,font=normalsize,labelfont=sf,textfont=sf]{subfig}
\usepackage{textcomp}
\usepackage{stfloats}
\usepackage{url}
\usepackage{verbatim}
\usepackage[utf8]{inputenc}
\usepackage[dvipsnames]{xcolor}
\usepackage{tabularray}
\usepackage[backend=biber,style=numeric,sorting=none]{biblatex}
\usepackage{graphicx}
\usepackage{rotating}
\usepackage{mdframed,lipsum}
\usepackage{enumitem}
\usepackage{soul,color}
\usepackage{booktabs}
\usepackage{xurl}
\usepackage{pdflscape}
\usepackage{longtable}
\usepackage{caption}
\usepackage{tabularx,booktabs,array,ragged2e}
\newcolumntype{Y}{>{\RaggedRight\arraybackslash\hspace{0pt}}X} 
\newcolumntype{C}[1]{>{\Centering\arraybackslash}p{#1}}   
\usepackage{hyperref}
\usepackage{comment}
\usepackage{ragged2e}
\usepackage{comment}
\usepackage{multirow}
\usepackage{pifont} 
\usepackage{xcolor} 
\usepackage{wasysym} 
\usepackage{adjustbox} 
\usepackage{tabularx,booktabs,array,ragged2e,multirow,pifont}
\usepackage{graphicx}
\usepackage{tikz}
\usetikzlibrary{trees,positioning,shapes.geometric,arrows.meta,
                  decorations.pathreplacing,calc,backgrounds,fit}
\usepackage{forest}          
\usepackage{makecell}       
\definecolor{adversary}{HTML}{C0392B}   
\definecolor{governance}{HTML}{1A6EA8}  
\definecolor{impact}{HTML}{27AE60}      
\definecolor{dimcol}{HTML}{5D6D7E}      

\newcommand{\Full}{\CIRCLE}
\newcommand{\Partial}{\RIGHTcircle}
\newcommand{\None}{--}
\AtBeginDocument{%
  }

\setcopyright{acmlicensed}
\copyrightyear{2018}
\acmYear{2018}
\acmDOI{XXXXXXX.XXXXXXX}

\acmJournal{JDS}
\acmVolume{37}
\acmNumber{4}
\acmArticle{111}
\acmMonth{8}

\begin{document}

\title{ICS Cybersecurity Datasets: A Systematic Meta-Review of Coverage, Evaluation Practice, and Structural Gaps}

\author{Konstantinos E. Kampourakis}
\email{konstantinos.kampourakis@ntnu.no}
\orcid{0009-0000-8883-0735}
\authornote{Corresponding author.}
\author{Vyron Kampourakis}
\email{vyron.kampourakis@ntnu.no}
\orcid{0000-0003-4492-5104}
\affiliation{%
  \institution{Norwegian University of Science and Technology}
  \city{Gjøvik}
  \country{Norway}
}

\author{Georgios Kambourakis}
\email{gkamb@aegean.gr}
\orcid{0000-0001-6348-5031}
\affiliation{%
  \institution{University of the Aegean}
  \city{Samos}
  \country{Greece}
}

\author{Sokratis Katsikas}
\email{sokratis.katsikas@ntnu.no}
\orcid{0000-0003-2966-9683}
\affiliation{%
  \institution{Norwegian University of Science and Technology}
  \city{Gjøvik}
  \country{Norway}
}

\author{Stefanos Gritzalis}
\email{sgritz@unipi.gr}
\orcid{0000-0002-8037-2191}
\affiliation{%
 \institution{University of Piraeus}
 \city{Athens}
 \country{Greece}}
 
\author{Mario Rodríguez-Béjar}
\email{mario.rodriguezb1@um.es}
\affiliation{%
 \institution{Universidad de Murcia}
 \city{Murcia}
 \country{Spain}}
 
\author{José Luis Hernández-Ramos}
\email{jluis.hernandez@um.es}
\orcid{0000-0001-7697-116X}
\affiliation{%
 \institution{Universidad de Murcia}
 \city{Murcia}
 \country{Spain}}

\renewcommand{\shortauthors}{Kampourakis et al.}
\acmArticleType{Review}
\acmCodeLink{https://github.com/borisveytsman/acmart}
\acmDataLink{htps://zenodo.org/link}
\acmContributions{BT and GKMT designed the study; LT, VB, and AP
  conducted the experiments, BR, HC, CP and JS analyzed the results,
  JPK developed analytical predictions, all authors participated in
  writing the manuscript.}

\begin{abstract}
Intrusion detection research in Industrial Control Systems (ICS) heavily depends on public datasets, yet no prior work has systematically assessed whether the collective dataset corpus supports current evaluation claims. This paper addresses this gap through a meta-review of 18 studies between 2019 and 2026, from which 83 ICS, or ICS directly related, cybersecurity datasets are identified, harmonised, and characterised using a unified five-dimensional taxonomy. The taxonomy reveals that the corpus is structurally skewed: 85.5\% of datasets concentrate on late-stage OT~Disruption tactics, cross-stage IT/OT progression sequences are present in only 8.4\% of cases, field-device evidence at Level~0 of the Purdue hierarchy is effectively absent, and operationally sourced data accounts for only 15.7\% of the collection. A parallel audit of evaluation practices shows that zero report streaming evaluation, fewer than half apply disciplined train/test partitioning, and only two satisfy reproducibility requirements. Furthermore, a taxonomy–evaluation coupling analysis shows that dataset imbalances constrain the scope and feasibility of several evaluation practices. Based on these findings, we identify three structural imbalances: i) architectural shallowness, ii) progression compression, and iii) cross-domain substitution, and derive a coordinated research agenda which covers cross-stage corpus construction, temporally structured benchmarking, event-level label standards, and governance frameworks for operational data sharing. 
\end{abstract}

\maketitle

\section{Introduction}
\label{S:Intro}

Industrial Control Systems (ICS) are essential for Critical Infrastructure (CI) operations, making their protection a matter of both operational continuity and societal resilience. The close integration of digital platforms and Industrial Internet of Things (IIoT) technologies with Information Technology (IT) systems, further exposes ICS environments to cyber risks that extend beyond conventional enterprise security concerns. Consequently, the cybersecurity of Operational Technology (OT) has emerged as a major research priority. For example, the landmark Stuxnet attack~\cite{falliere2011w32} marked a pivotal moment in demonstrating the vulnerability of industrial environments to sophisticated state-sponsored cyber operations, while more recent campaigns such as Sandworm~\cite{Sandstorm} underscored the persistence and evolution of such threats.

Publicly accessible datasets are indispensable in this research area. They enable anomaly detection, attack classification, benchmark comparison, and reproducible evaluation of security methods, while providing a shared basis for assessing Machine Learning (ML) and other data-driven approaches. Public datasets provide a common experimental substrate for the community~\cite{kenyon2020public}, in contrast to proprietary industrial data, which are often inaccessible due to their constraints. As a result, their role in accelerating methodological development and supporting the transition toward practical and deployable defense mechanisms, is crucial.

Despite their importance, the ICS dataset ecosystem remains fragmented. Existing datasets are unevenly distributed across industrial sectors and data modalities, and they differ substantially in attack representation, labeling conventions, and documentation quality~\cite{conti2021survey, dobler2025systematic}. In addition, evaluation practices vary widely across studies, where metrics, validation procedures, and experimental splits are used inconsistently~\cite{koay2023machine, martins2024comparative}. This further complicates fair comparison. Perhaps most critically, evidence for cross-dataset generalization remains limited, which raises doubts about the transferability of models and techniques. These limitations collectively hinder reproducibility, comparability, and the formation of a coherent empirical foundation for ICS cybersecurity research~\cite{conti2021survey}.

Several attempts to organize this landscape have been made. Nevertheless, some surveys mostly focus on cataloging publicly available datasets and their technical characteristics~\cite{conti2021survey,dobler2025systematic,holdbrook2024network,ekisa2024virtual}, while others emphasize intrusion detection applications and ML-based analyses~\cite{mubarak2021anomaly,rakas2020review,alanazi2023scada}. Taken together, these studies confirm the central role of datasets in ICS cybersecurity. However, they also reveal persistent gaps, including inconsistent taxonomies, incomplete sectoral coverage, methodological fragmentation, and limited attention to generalization and reproducibility.

This meta-review aims to address these shortcomings. The main goal is to synthesize the recent survey literature into a unified and structured account of the ICS cybersecurity dataset landscape and integrate evidence across sectors, modalities, and review perspectives. Specifically, this includes dataset-centric, IDS-centric, testbed-oriented, and domain-focused studies. Also, the work at hand reconciles fragmented characterization schemes into a standard-anchored analytical framework, examines the limited evidence on cross-dataset evaluation and reproducibility, and consolidates the research gaps and recommendations scattered across prior surveys into a coherent agenda for future work. In doing so, the review clarifies the current state of public ICS dataset research and supports a more realistic, reproducible, and generalizable basis for cybersecurity evaluation in critical infrastructure environments. Specifically, this work makes the following contributions:

\begin{enumerate}
  \item \textbf{Systematic survey-level evidence synthesis.} We conduct a PRISMA-guided meta-review of recent peer-reviewed survey and review studies on public ICS, SCADA, OT, and IIoT cybersecurity datasets. 

  \item \textbf{Unified taxonomy grounded in ICS standards.} We consolidate fragmented prior characterization schemes into a five-dimensional taxonomy that captures adversary lifecycle coverage, attack progression depth, architectural evidence layer, dataset provenance and fidelity, and ground-truth quality. 

  \item \textbf{Corpus-level characterization of identified datasets.} We extract and synthesize the datasets identified across the included studies and apply the unified taxonomy to expose dominant structural patterns.

  \item \textbf{Reproducibility and evaluation analysis.} We examine how prior surveys report labeling practices, validation protocols, performance metrics, dataset splits, and artifact availability, with emphasis on recurring weaknesses.

  \item \textbf{Gap analysis and forward-looking agenda.} We distill the principal limitations identified across the survey literature, and translate them into actionable recommendations for dataset creators, benchmark curators, and future evaluators.
\end{enumerate}

The remainder of this paper is organised as follows. Section~\ref{S:meth} presents the systematic review methodology. Section~\ref{S:overview} characterises the evidence base. Section~\ref{S:Taxonomy} surveys the taxonomy families identified in prior work and establishes the motivating case for a unified framework. Section~\ref{S:unified_taxonomy} introduces the proposed five-dimensional taxonomy. Section~\ref{S:comparative_dataset_analysis} provides a comparative analysis of the dataset corpus. Section~\ref{S:Evaluation} audits the evaluation practices reported across the included studies. Section~\ref{S:RoadAhead} derives a forward-looking agenda. The last section concludes the paper.

\section{Methodology}
\label{S:meth}

\begin{figure*}[htbp!]
    \centering
    \includegraphics[width=\linewidth]{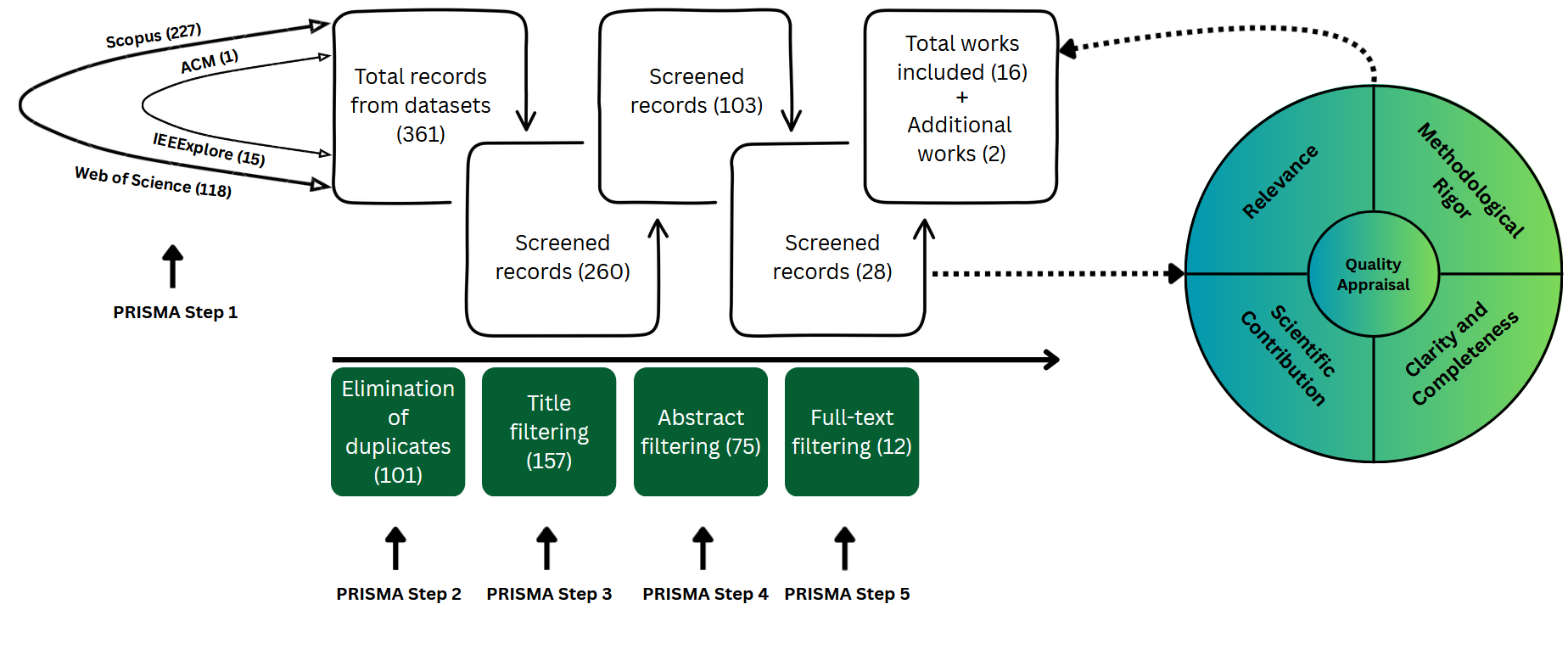}
    \caption{Overview of the identification, screening, eligibility, and inclusion process used in this meta-review.}
    \label{F:lit:screen}
\end{figure*}

As discussed previously, the present work contributes a meta-review of research on ICS cybersecurity datasets. Accordingly, the primary evidence unit for identification and screening is the included study, while the datasets extracted from those studies constitute the secondary analytical unit for corpus-level characterization. Reviews of testbeds and related experimental environments, such as provenance, fidelity, architectural scope, protocol coverage, attack execution, labeling, or reproducibility are included only when they provide evidence concerning datasets generated from those environments or characteristics that directly affect dataset interpretation. To guide the collection and screening of papers, several established approaches for conducting systematic literature reviews were examined~\cite{Pagen71,bell2022business,creswell2016qualitative,lacerda2018systematic}. In this respect, we adopted the systematic review framework proposed by~\cite{fink2019conducting,okoli2010guide}, as it offers transparent and reproducible procedures for synthesizing both quantitative and qualitative evidence. Furthermore, this review adheres to the Preferred Reporting Items for Systematic Reviews and Meta-Analyses (PRISMA) statement~\cite{Pagen71,fink2019conducting}, which has been widely applied in recent reviews in related cybersecurity and cyber-physical systems domains~\cite{kannelonning2023systematic,kamp2023}. Based on the PRISMA workflow, the inclusion and exclusion criteria, and the quality appraisal procedure, the literature screening process is summarized in Figure~\ref{F:lit:screen}.

\noindent \textbf{Step I -- Research Questions:} The objective of this meta-review is to identify, analyze, and synthesize survey and review studies that address the use of datasets in ICS cybersecurity research. More specifically, we seek to consolidate knowledge on the types of datasets used, the security tasks they support, the industrial domains they represent, and the gaps that limit progress in the field. Accordingly, the following research questions (RQs) were established:

\begin{itemize}
    \item \textbf{RQ1:} What dataset types, benchmarks, generation environments, and survey perspectives recur most frequently in the ICS cybersecurity review literature? This question identifies the experimental resources and review emphases that dominate the literature, including dataset-centric, IDS-centric, and testbed-oriented studies. Note that testbeds are considered only where they serve as dataset-generation or characterization environments and provide evidence regarding dataset provenance, fidelity, architectural scope, or reproducibility.

    \item \textbf{RQ2:} Which adversary lifecycle stages, attack progression depths, system and evidence layers, provenance classes, and detection-evidence characteristics are most commonly represented in the identified dataset corpus? This question characterizes the technical and evidential scope of the corpus, with particular attention to disruption-heavy attack coverage, limited cross-stage progression, uneven architectural visibility, operational realism, and label quality.

    \item \textbf{RQ3:} What methodological limitations, dataset deficiencies, and evidence gaps are identified by prior surveys and reviews? This question examines frequently observed weaknesses in realism, labeling granularity, provenance reporting, reproducibility, and cross-dataset generalization.

    \item \textbf{RQ4:} What future research directions are recommended to improve dataset realism, operational relevance, and evaluation credibility for ICS cybersecurity? This question synthesizes the forward-looking recommendations reported in the included studies and translates them into an actionable research agenda.
\end{itemize}

\noindent \textbf{Step II -- Bibliographic Sources:} The review was conducted using four major academic databases: IEEE Xplore, Scopus, Web of Science, and ACM Digital Library, as shown in Figure~\ref{F:lit:screen}. These sources were selected because they provide broad coverage of computer science, engineering, and security research, and are widely recognized as suitable repositories for systematic reviews in this area~\cite{silvasystematic}. In this respect, using multiple databases helped reduce retrieval bias and improved the likelihood of capturing relevant studies across publication venues.

\noindent \textbf{Step III -- Search Terms:} The search strategy was designed around four conceptual groups of terms. The first group targeted the industrial context of interest, namely ICS, SCADA, OT, IIoT, and related CI environments. The second group focused on dataset-related terms such as dataset, benchmark, testbed, and data collection. The third group targeted cybersecurity-related concepts, including cybersecurity, intrusion detection, anomaly detection, attack, and IDS. The fourth group captured review-oriented publications such as surveys, reviews, meta-analyses, systematic reviews, and literature reviews. The logical combinations of these terms are provided in Table~\ref{tab:terms}. The complete database-specific search strings are provided in Appendix~\ref{app:queries}. Applying this strategy across the selected databases resulted to an initial corpus of 361 records.

\begin{table*}[htbp!]
\centering
\caption{Search string groups and their logical combinations.}
\label{tab:terms}
\resizebox{\textwidth}{!}{
\begin{tabular}{p{6cm}lll}
\toprule
\textbf{Group 1: Domain Terms} & \textbf{Group 2: Dataset Terms} & \textbf{Group 3: Security Terms} & \textbf{Group 4: Review-Type Terms} \\
\midrule
Industrial Control System, ICS, SCADA, OT, IIoT, Operational Technology, Critical Infrastructure
&
\begin{tabular}[c]{@{}l@{}}dataset\\ benchmark\\ testbed\\ data collection\end{tabular}
&
\begin{tabular}[c]{@{}l@{}}cybersecurity\\ intrusion detection\\ anomaly detection\\ attack\\ IDS\end{tabular}
&
\begin{tabular}[c]{@{}l@{}}survey\\ review\\ meta-analysis\\ systematic review\\ literature review\end{tabular}
\\
\midrule
\multicolumn{4}{c}{\textbf{Search logic:} Group 1 \textbf{AND} Group 2 \textbf{AND} Group 3 \textbf{AND} Group 4} \\
\bottomrule
\end{tabular}}
\end{table*}

\noindent \textbf{Step IV -- Practical Screening:} After retrieval, a practical screening phase was conducted to ensure that only relevant and high-quality studies were included in the review. In line with PRISMA and to reduce selection bias, the screening process was performed systematically using predefined inclusion and exclusion criteria, summarized in Table~\ref{tab:criteria}. Note that 2019–-2026 (IC3 in table~\ref{tab:criteria}) publication window was selected deliberately to capture contemporary developments in ICS cybersecurity datasets and their use in modern data-driven intrusion-detection research. The screening procedure was executed in four stages: first, duplicate records were removed; second, titles were screened; third, abstracts were assessed for relevance and eligibility; and finally, full-text articles were examined to verify compliance with the inclusion criteria.

The screening and eligibility assessment were conducted by members of the author team. Where inclusion or exclusion judgments differed, the involved reviewers discussed the case until consensus was reached. This cross-review procedure reduced the risk of individual reviewer-selection bias and prevented inclusion decisions from relying on a single assessor. Ultimately, only studies that explicitly addressed ICS cybersecurity datasets, benchmarks, testbeds, or dataset-related survey evidence were retained for detailed analysis. For example, works like the ones in~\cite{kheddar2024reinforcement,sheng2025network} are deliberately excluded, as they do not review or compare public ICS cybersecurity datasets. Instead, they focus on RL-based IDS across generic IT/IoT networks, IIoT device fingerprinting, or virtual/DT testbed architectures. To further reduce subjectivity and other potential sources of bias, predefined inclusion and exclusion criteria were applied consistently throughout the screening process. Moreover, a quality appraisal was applied to each study based on methodological clarity, relevance to the research questions, and overall contribution to the field.

\begin{table*}[!t]
\renewcommand{\arraystretch}{1.25}
\caption{Inclusion and Exclusion Criteria for Full-Text Eligibility}
\label{tab:criteria}
\centering
\small
\begin{tabular}{p{0.48\textwidth} p{0.48\textwidth}}
\hline
\textbf{Inclusion Criteria} & \textbf{Exclusion Criteria} \\
\hline
\textbf{IC1}: Reviews, surveys, or meta-analyses of publicly available ICS, SCADA, OT, or IIoT cybersecurity datasets &
\textbf{EC1}: Focuses exclusively on dataset \emph{creation} without reviewing or comparing existing datasets \\[2pt]

\textbf{IC2}: Covers at least two distinct public datasets and provides a comparative or taxonomic analysis &
\textbf{EC2}: Exclusively targets generic IT network datasets (e.g., KDD~Cup~99, NSL-KDD, CICIDS) with no ICS-specific content \\[2pt]

\textbf{IC3}: Published in a peer-reviewed venue (journal, conference, or workshop with proceedings) between 2019 and 2026 &
\textbf{EC3}: Focuses on ML, DL, or RL-based IDS, device fingerprinting, or digital-twin testbed design without dataset review \\[2pt]

\textbf{IC4}: Provides explicit discussion of dataset properties (modality, attack coverage, labeling, or evaluation methodology) &
\textbf{EC4}: Duplicate publication of a paper already included (e.g., extended journal version supersedes conference version) \\
\hline
\end{tabular}
\end{table*}

The systematic database screening resulted in 16 studies that satisfied the review-oriented eligibility criteria. To broaden coverage of directly relevant dataset-characterization evidence, backward and forward citation tracing was subsequently performed on the retained studies. This process identified two additional closely related works~\cite{gomez2019generation} and~\cite{mubarak2021anomaly}. Although these studies are not conventional survey articles, they provide structured comparative or taxonomic evidence directly relevant to the objectives of this meta-review. Thus, the final corpus comprises 18 studies: 16 surveys or reviews identified through the systematic screening procedure and two supplementary comparative studies identified through citation tracing. Collectively, these studies provide relevant evidence on ICS cybersecurity datasets, including dataset taxonomies, benchmark analyses, testbed reviews, comparative evaluations, and discussions of dataset quality and applicability. The retained studies form the basis for the synthesis and discussion presented in the subsequent sections.

\subsection{Threats to Validity}

Despite closely abiding by the PRISMA guidelines, several threats to validity remain. First, selection and reviewer bias were reduced through predefined eligibility criteria, multi-stage screening, cross-review of inclusion decisions, and consensus-based resolution of disagreements. Moreover, retrieval bias was mitigated by searching four major bibliographic databases and by applying equivalent conceptual search blocks across database-specific query syntaxes. Nevertheless, terminology variation in the ICS, OT, SCADA, and IIoT literature may have caused relevant studies to remain undetected.

Second, publication and reporting bias cannot be fully eliminated because the review is limited to peer-reviewed literature and relies on the information reported by the included studies. In other words, as this study synthesizes survey-level evidence, omissions, inconsistent reporting, or classification decisions in the primary surveys may propagate into the present analysis. Similarly, datasets that have not been covered by the included studies may be absent from the resulting corpus.

Third, a threat concerns classification and extraction subjectivity. The harmonization of heterogeneous datasets into the taxonomy later presented in Section~\ref{S:unified_taxonomy} necessarily involves interpretive decisions, particularly when original publications provide incomplete information about attack progression, architectural scope, provenance, or labeling. To reduce this risk, each taxonomy dimension was coded using predefined decision rules derived from the corresponding reference framework. MITRE ATT\&CK for ICS~\cite{mitre_ics} was used to classify adversarial tactic coverage, the ICS Cyber Kill Chain~\cite{sans_icskc} to distinguish Stage~1, Stage~2, and cross-stage progression, IEC~62443~\cite{iec62443} together with the Purdue Reference Model to characterize architectural evidence scope, NIST SP~800-82r3~\cite{nist80082} to classify operational provenance and environment type, and the NIST CSF~\cite{nist_csf_2024} to characterize labeling and detection-evidence quality. The same decision rules were applied consistently across the corpus. Nevertheless, some residual classification uncertainty cannot be completely excluded.

Finally, the quantitative distributions later reported throughout the paper should be interpreted as descriptive properties of the dataset corpus identified through the included surveys and not as exhaustive estimates of the complete space of ICS cybersecurity datasets. However, these limitations do not invalidate the comparative analysis, but they constrain the generality of the resulting corpus-level conclusions.

\section{Evidence Base and Prior Surveys}
\label{S:overview}

This section first characterizes the overall evidence base by examining when the included studies were published, how they are thematically distributed, and how broadly they cover the relevant analytical dimensions. It then synthesizes it according to its primary emphasis, distinguishing between dataset-centric, IDS- and method-centric, and testbed- or domain-focused reviews. Together, the following subsections establish the scope, diversity, and main limitations of the existing survey landscape on ICS datasets.

\subsection{Corpus Overview}

Here, we provide a descriptive overview of the 18 studies retained for the meta-review. Specifically, we characterize the evidence base before the comparative synthesis, through the examination of the temporal distribution, thematic emphasis, and analytical breadth. Although the included studies share a common focus on ICS cybersecurity datasets, they vary substantially in scope, level of detail, and analytical emphasis, as summarized in Table~\ref{tab:included_studies_overview}.

Namely, some focus primarily on cataloging publicly available datasets and testbeds, while others examine the datasets indirectly through the lens of IDS methods, protocol security, or ML-based evaluation. As a result, the evidence base is not methodologically uniform, but rather composed of partially overlapping survey perspectives that collectively focus on different aspects of the dataset ecosystem. Moreover, this section analyzes the 18 selected studies, providing a summary of the collected surveys on ICS/SCADA/IIoT datasets for intrusion detection, and filtering out key points per work, such as the categorization of dataset/attack taxonomies, feature extraction and data modalities, employed public testbeds and datasets, attack types, performance and evaluation protocols, and research gaps.

\begin{table*}[htbp!]
\centering
\caption{A reverse chronological overview of the studies included in the final meta-review.}
\label{tab:included_studies_overview}
\resizebox{\textwidth}{!}{%
\begin{tabular}{p{0.24\textwidth}p{0.08\textwidth}p{0.3\textwidth}p{0.5\textwidth}}
\hline
\textbf{Study} & \textbf{Year} & \textbf{Scope} & \textbf{Principal contribution} \\
\hline
Raza et al.~\cite{raza2026deeplearning} & 2026 & Deep learning-based intrusion detection and datasets for IIoT security. & IIoT cybersecurity challenges, IIoT datasets, and recent DL-based IDS in industrial environments. \\
Sanjalawe et al.~\cite{Sanjalawe2026} & 2026 & AI-based intrusion detection in IIoT environments. & IIoT-focused taxonomy of IDS strategies and AI methods, deployment considerations across edge/fog/cloud, explainability, adversarial robustness, and benchmark limitations. \\
Afaq et al.~\cite{afaq2025publicdatasets} & 2025 & Public datasets for ML-based intrusion detection in IoT and OT networks. & Publicly available IoT/OT intrusion datasets, protocols, attacks, features, and metadata, and gaps in realism, balance, and protocols. \\
Dobler et al.~\cite{dobler2025systematic} & 2025 & Industrial network traffic datasets. & 32 open-access IT, IoT, IIoT, OT, and ICS malicious traffic datasets, 97 attack types to the Cyber Kill Chain, and dataset complexity and feature characteristics. \\
Martins et al.~\cite{martins2024comparative} & 2024 & Cybersecurity datasets in ICS. & OT/ICS IDS datasets, composition, attack coverage, imbalance, and limitations, lack of multistage attacks. \\
Muhammad et al.~\cite{muhammad2024smartgrid} & 2024 & Cybersecurity in smart grids protocols and datasets. & Smart grid communication protocols and standards, four smart grid-related datasets, dataset completeness, and research directions. \\
Elouardi et al.~\cite{elouardi2024hybridcnnllm} & 2024 & Hybrid CNN and LLM approaches for IDS using IoT datasets. & AI-based IDS methods, recent IoT datasets, and hybrid CNN/LSTM and BERT-based approaches for IDS in IoT settings. \\
Ekisa et al.~\cite{ekisa2024virtual} & 2024 & Virtual ICS Testbeds. & 21 virtual ICS testbeds, virtualized components, protocols, attack scenarios, virtualization technologies, and reproducibility, and gaps in open access and modern container-based approaches. \\
Alanazi et al.~\cite{alanazi2023scada} & 2023 & SCADA-focused vulnerabilities, attacks, IDS, and testbeds. & Taxonomies of SCADA vulnerabilities, attack types, IDS categories, and testbeds, SCADA protocols, security requirements, and open issues. \\
Alex et al.~\cite{alex2023iotdatasets} & 2023 & IoT security datasets, taxonomy, classification, and ML mechanisms. & 44 IoT/IIoT datasets, dataset taxonomies and classification schemes, and ML methods for dataset evaluation. \\
Alani et al.~\cite{alani2023survey} & 2023 & Smart grid IDS datasets. & Five smart grid datasets, protocol coverage, samples, attacks, class balance, and suitability for ML-based IDS. \\
Koay et al.~\cite{koay2023machine} & 2023 & Machine learning in ICS security. & ICS vulnerabilities, ML-based attack detection methods and datasets, and limitations, dataset quality, and adversarial risk. \\
Yadav and Paul~\cite{yadav2021architecture} & 2021 & SCADA system architecture, communication protocols, attacks, IDS, testbeds, and IoT-based SCADA. & Taxonomy linking SCADA architecture, protocols, attack types, IDS techniques, and testbed approaches, open research gaps in end-to-end SCADA security. \\
Conti et al.~\cite{conti2021survey} & 2021 & ICS testbeds and datasets for security. & Overview of ICS architectures, industrial protocols, testbeds, and datasets, and best-performing IDS baselines, design guidelines and good practices for future testbed/dataset development. \\
Mubarak et al.~\cite{mubarak2021anomaly} & 2021 & Anomaly detection in ICS datasets with machine learning algorithms. & Public SCADA/ICS datasets, ML classifiers for anomaly detection, profiling network traffic, build baseline-driven behavior analysis for ICS security. \\
Rakas et al.~\cite{rakas2020review} & 2020 & Network-based SCADA IDS. & 26 SCADA NIDS by detection methodology, protected protocols, implementation tools, test environments, and performance evaluation, and key research gaps and future directions. \\
Lin et al.~\cite{lin2019using} & 2019 & Datasets from ICS for cybersecurity research and education. & Six public ICS datasets, value of real ICS data for both research and teaching. \\
Gómez et al.~\cite{gomez2019generation} & 2019 & Anomaly detection datasets in ICS. & Four-step methodology for generating realistic ICS anomaly datasets, Electra dataset, supervised and semi-supervised anomaly detection models on it. \\
\hline
\end{tabular}}
\end{table*}

\begin{figure}[htbp!]
    \centering
    \includegraphics[width=0.5\columnwidth]{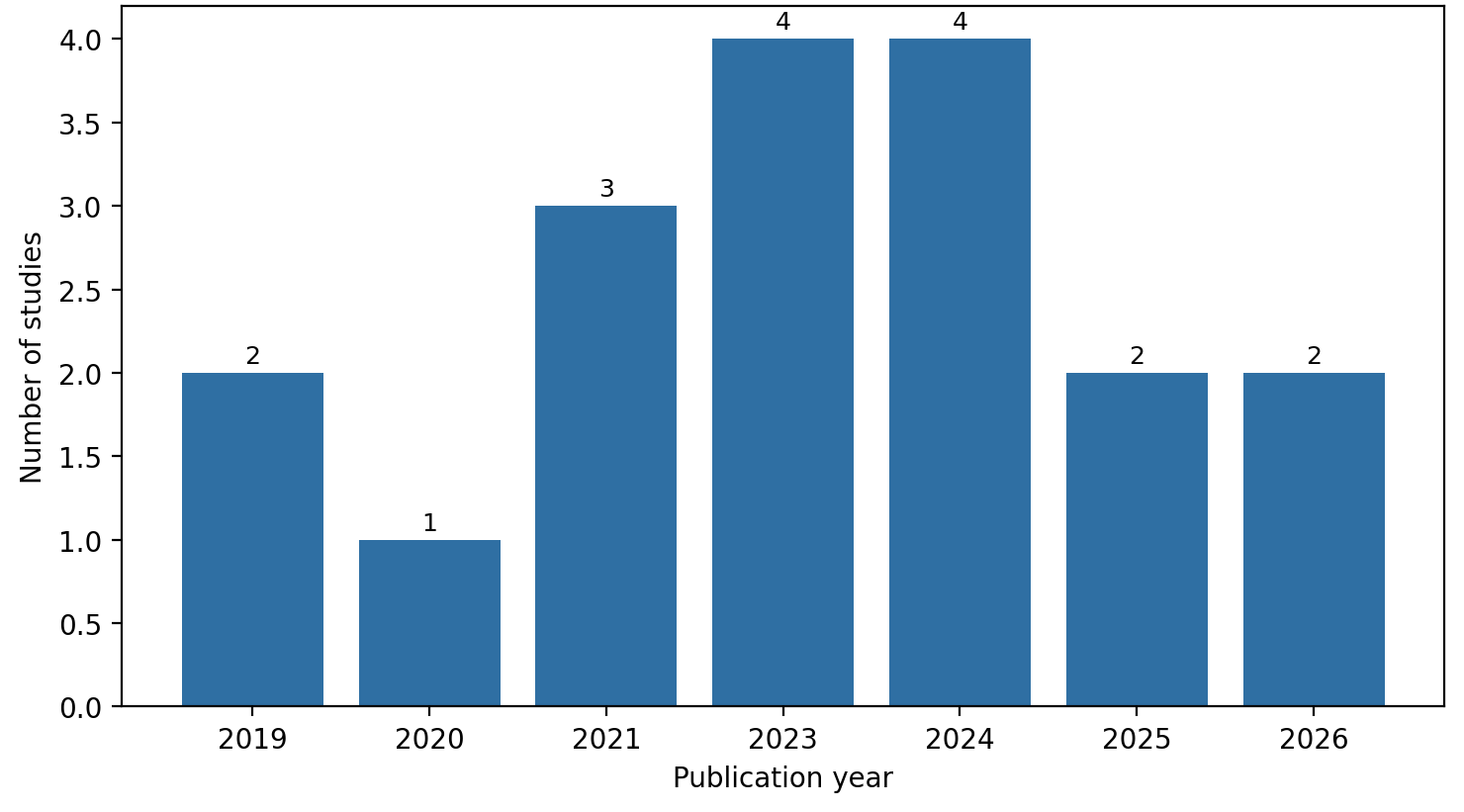}
    \caption{Temporal distribution of the studies included in the meta-review.}
    \label{fig:publication_timeline}
\end{figure}

From a temporal perspective, the literature reflects a marked increase in attention to ICS cybersecurity datasets in recent years. The included studies are concentrated from 2019 onward, indicating that systematic reflection on public ICS datasets is a relatively recent development compared with the longer history of ICS security research more broadly, as depicted in Figure~\ref{fig:publication_timeline}. This increase is likely associated with the growing reliance on data-driven IDS approaches, the wider availability of public benchmarks, and a parallel recognition that dataset realism, diversity, and documentation quality have become central concerns for credible evaluation.

\begin{figure*}[htbp!]
    \centering
    \includegraphics[width=\linewidth]{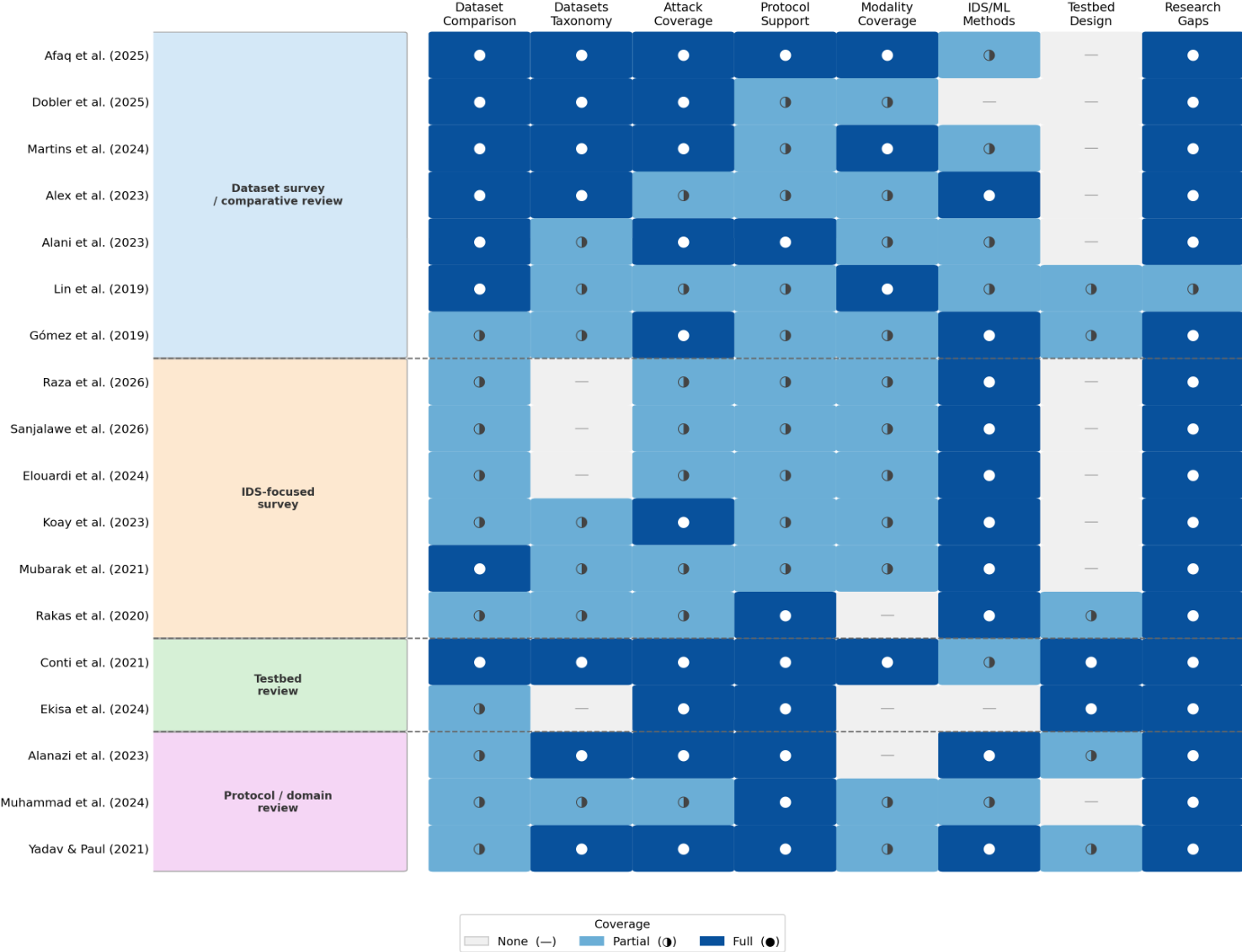}
    \caption{Analytical breadth of the included studies across key thematic dimensions.}
    \label{fig:evidence_breadth}
\end{figure*}

A second notable feature of the evidence base is its uneven thematic coverage. Not all included studies examine datasets in the same way. Some provide explicit comparative analyses of publicly available datasets and discuss features like attack diversity, protocol support, class balance, or modality~\cite{afaq2025publicdatasets,dobler2025systematic,martins2024comparative,alex2023iotdatasets,alani2023survey,lin2019using,gomez2019generation}. Others focus more broadly on IDS in ICS and treat datasets primarily as supporting artifacts~\cite{raza2026deeplearning,Sanjalawe2026,elouardi2024hybridcnnllm,koay2023machine,mubarak2021anomaly,rakas2020review}. Still others emphasize testbeds and simulation environments, treating datasets as outputs of those experimental infrastructures~\cite{conti2021survey,ekisa2024virtual}. Consequently, the literature includes both dataset-centric and method-centric perspectives, and this distinction is important for interpreting the depth and type of evidence extracted from each study, as shown in Figure~\ref{fig:evidence_breadth}.

The analytical breadth of the included studies is similarly uneven, as illustrated in Figure~\ref{fig:evidence_breadth}. Only a subset of the surveyed works provides a clear taxonomy for organizing datasets, attacks, or evaluation criteria~\cite{dobler2025systematic,martins2024comparative,alex2023iotdatasets,conti2021survey,alanazi2023scada,yadav2021architecture}. Likewise, only some studies conduct systematic side-by-side dataset comparisons~\cite{afaq2025publicdatasets,dobler2025systematic,martins2024comparative,alex2023iotdatasets,alani2023survey,lin2019using,mubarak2021anomaly,conti2021survey}, while others provide descriptive summaries or selective exemplars~\cite{gomez2019generation,raza2026deeplearning,Sanjalawe2026,elouardi2024hybridcnnllm,koay2023machine,rakas2020review,ekisa2024virtual,alanazi2023scada,muhammad2024smartgrid,yadav2021architecture}. In contrast, the discussion of research gaps is more widespread: all studies identify at least some recurring limitations, with respect to realism, class imbalance, limited attack diversity, and insufficient evaluation consistency. This asymmetry suggests that the field has matured more quickly in diagnosing problems than in establishing standardized analytical frameworks for comparing datasets.

An important observation from Figure~\ref{fig:evidence_breadth} concerns the balance between breadth and depth in the surveyed literature. Broad surveys often cover a larger number of datasets or application contexts. However, they may do so at the expense of detailed discussion of labeling quality, data provenance, or evaluation practice~\cite{conti2021survey,koay2023machine}. Conversely, narrower reviews sometimes provide richer analysis of a smaller set of datasets or sectors, especially in relation to specific domains such as SCADA, smart grids, or IIoT~\cite{alanazi2023scada,muhammad2024smartgrid,yadav2021architecture,alex2023iotdatasets}. This trade-off explains why a meta-review is important: no single prior survey simultaneously offers broad coverage, consistent taxonomy, and deep methodological critique across the full range of publicly available ICS cybersecurity datasets.

Overall, the evidence base is quite valuable, but still, structurally uneven. On one hand, it provides sufficient material to identify recurring datasets, dominant sectors, and shared concerns. On the other hand, it exhibits inconsistency in terminology, analytical rigor, and thematic focus. These characteristics justify the need for a synthesis framework. Therefore, the subsequent sections move beyond description and examine how prior studies classify datasets and attacks, laying the groundwork for a unified taxonomy.

\subsection{Prior Survey Literature}
\label{SS:Related}

Based on their primary analytical emphasis, the reviewed studies were organized into three broad categories: dataset-centric surveys, IDS- and method-centric surveys, and testbed- and domain-focused surveys. This grouping provides a structured basis for comparing how prior work approaches the ICS cybersecurity dataset landscape and highlights the distinct, yet partially overlapping, perspectives represented in the literature.

\subsubsection{Dataset-Centric Surveys}

Several dataset-centric studies compare publicly available IoT, IIoT, OT, and ICS datasets according to their suitability for ML-based intrusion detection. The authors in~\cite{afaq2025publicdatasets} compare 14 datasets using protocol and attack coverage, metadata, sample size, complexity, labeling quality, and scalability. They also  identify limited protocol diversity, weak multi-stage APT coverage, class imbalance, and simplified attack scripts. Moreover, the work in~\cite{dobler2025systematic} systematically reviews open ICS/IIoT network-traffic datasets and examine attack coverage, labeling quality, class imbalance, and feature utility, while also identifying temporal artifacts and address-based labeling as recurring issues. For attack characterization, it uses Cyber Kill Chain stages, a framework whose technical structure and attack-stage interpretation are discussed in~\cite{yadav2015technical}. The study in~\cite{martins2024comparative} compares OT and ICS datasets by attack coverage, feature composition, labeling consistency, and class balance, emphasizing the dominance of single-stage attacks, the lack of multistage or hybrid threat sequences, data scarcity, and the absence of standardized evaluation criteria.

A broader IoT/IIoT perspective is provided in~\cite{alex2023iotdatasets}, which classifies 44 datasets according to availability, format, size, feature characteristics, dataset type, test-bed setup, and attack coverage, highlighting limited realistic multi-layer settings, insufficient documentation, narrow protocol coverage, and scarce production-like test-beds. In the smart-grid domain, the authors in~\cite{alani2023survey} compare public intrusion-detection datasets by protocol coverage, feature dimensionality, sample size, attack diversity, labeling strategy, and balance, distinguishing legacy generic datasets from domain-specific datasets that capture authentic SCADA communications. The work in~\cite{lin2019using} reviews six public ICS datasets from operational and simulated environments, including SWaT, WADI, and BATADAL, and discusses their data modalities, attack coverage, accessibility, and use in reproducible research and cybersecurity education.

The study in~\cite{gomez2019generation} addresses the dataset-generation perspective by proposing a four-stage methodology comprising attack selection, attack deployment, traffic capture, and feature computation. The methodology is demonstrated through the Electra dataset, representing an electric traction substation and containing normal and attack traffic generated through controlled cyberattack scenarios. The work also evaluates ML and DL anomaly-detection models on Electra and contributes primarily to realistic dataset generation, reproducibility, and validation, while underlining requirements for future ICS dataset development and evaluation.

Put together, dataset-centric surveys consistently identify dataset quality, realism, and representativeness as major constraints on credible ICS intrusion-detection evaluation. Regardless of differences in scope, they converge on recurring deficiencies and support the need for continuously updated, well-documented, and reproducibly prepared datasets.

\subsubsection{IDS- and Method-Centric Surveys}

IDS- and method-centric studies generally examine datasets through the requirements of ML- and AI-based intrusion detection. In~\cite{raza2026deeplearning}, the authors review deep-learning-based IDS for IIoT, classify IDS by scope and technique, and compare five widely used benchmarks, namely KDDCUP99, NSL-KDD, UNSW-NB15, SWaT, and TON\_IoT. The study highlights the lack of emerging attacks in older datasets, restricted accessibility or preprocessing burden in some industrial traces, and the need for continuously updated datasets that reflect realistic IIoT conditions. The survey in~\cite{Sanjalawe2026} provides a broader review of AI-based intrusion detection for IIoT through a multidimensional taxonomy covering learning paradigms, deployment architectures, and operational constraints. It also discusses SWaT, WADI, and TON\_IoT and identifies the need for realistic multimodal datasets, stronger privacy preservation, and more standardized evaluation practices.

The authors in~\cite{elouardi2024hybridcnnllm} focus on hybrid CNN- and LLM-oriented IDS approaches and compare recent models on TonIoT, CICIoT2023, and X-IIoTID. Their analysis highlights class imbalance in CICIoT2023, strong separability in X-IIoTID, minority-class weaknesses in TonIoT, and broader challenges involving scalability, real-time detection, and resource-aware IDS design. The work in~\cite{koay2023machine} reviews ML-based intrusion detection for ICS and compares datasets according to attack types, sector coverage, feature granularity, and labeling quality. It identifies overreliance on water and energy datasets, limited multistage or hybrid attack representation, inconsistent labeling, small-scale or simulated evaluation environments, and the lack of standardized evaluation for reproducible IDS benchmarking.

A more evaluation-oriented perspective is provided in~\cite{mubarak2021anomaly}, which benchmarks classical ML algorithms on public SCADA and PCAP-derived network data, reports accuracy, precision/recall, F1, and ROC measures, and applies information-gain feature ranking. The study argues that many public datasets are outdated or poorly suited to industrial scenarios and motivates the collection of more realistic OT traffic through sensors and testbeds. Finally, the work in~\cite{rakas2020review} reviews network-based SCADA IDS using a structured rubric covering detection methodology, protected protocols, implementation tools, test environment, and performance. It finds that only five of 26 studies use real SCADA network data, while software simulations dominate, and further highlights heterogeneous evaluation practices, limited reporting of timeliness and efficiency, and the need for common metrics, realistic testbeds, and access to real-network datasets.

Overall, IDS- and method-centric surveys primarily examine datasets as experimental substrates for comparing detection architectures, learning paradigms, and deployment strategies. Despite frequent reports of high predictive performance, the reviewed works consistently identify limited dataset realism, class imbalance, outdated attack scenarios, heterogeneous evaluation protocols, and insufficient consideration of latency, scalability, and resource constraints. This indicates that methodological advances should be assessed through standardized, reproducible, and operationally grounded evaluations and not through benchmark accuracy alone.

\subsubsection{Testbed- and Domain-Focused Surveys}

Recall from section~\ref{S:meth} that testbed-focused reviews are included in the present synthesis only where they contribute evidence about the datasets produced by those environments. Their relevance is therefore instrumental rather than independent, since testbed architecture, virtualization, process fidelity, protocol support, attack execution, and reproducibility directly shape dataset provenance and evidential scope. Accordingly, the present work does not attempt to provide a comparative taxonomy or evaluation of ICS testbeds themselves.

The contribution in~\cite{ekisa2024virtual} reviews 21 fully virtualised ICS testbeds and analyses their virtualised components, industrial sectors, protocols, reproduced attacks, virtualisation technologies, and reproducibility. It identifies a strong concentration in energy-related sectors, the dominance of Modbus, and poor reproducibility, with only three testbeds providing publicly available setups or code, while recommending greater openness, broader attack coverage, and improved scalability and fidelity. Alanazi et al.~\cite{alanazi2023scada} review SCADA vulnerabilities, attacks, datasets, and testbeds, highlighting issues such as duplicate or dated benchmark data and emphasizing reproducibility, domain fidelity, process simulation, and alignment with real OT behaviour. The survey in~\cite{conti2021survey} examines ICS security testbeds together with their released datasets, organizing them by sector, modality, attack coverage, and testbed type. It highlights uneven sector and attack coverage, inconsistent evaluation practices, and fidelity concerns, while recommending common metrics and multi-dataset validation.

The authors in~\cite{muhammad2024smartgrid} provide a protocol-centered review of smart-grid cybersecurity, covering major communication protocols and standards alongside four representative datasets. Their analysis shows that most of these datasets rely on simulated testbeds and PCAP/CSV captures, while only Electra is generated in a real-time scenario, motivating the need for modern, balanced, real-world traces. Finally, the work in~\cite{yadav2021architecture} reviews SCADA architectures, communication protocols, attacks, IDS techniques, and testbeds, and emphasizes their interdependence for end-to-end security. It identifies the scarcity of validated datasets, the common reliance on simulations and small laboratory traces, and the role of testbeds as practical data-generation environments when operational collection is infeasible.

Across testbed- and domain-focused surveys, the quality of cybersecurity evaluation is shown to depend heavily on the fidelity, openness, and architectural scope of the underlying experimental environment. The literature reveals persistent reliance on simulated or small-scale testbeds, narrow sector and protocol coverage, limited reproducibility, and scarce access to real operational traces. Therefore, to improve the evidence base more transparent and reusable testbeds, broader domain representation, realistic process behaviour, and closer alignment between generated data and actual OT conditions is required.

\section{Taxonomies in Related Work}
\label{S:Taxonomy}

\begin{figure*} [!ht]
    \centering
    \includegraphics[width=1\linewidth]{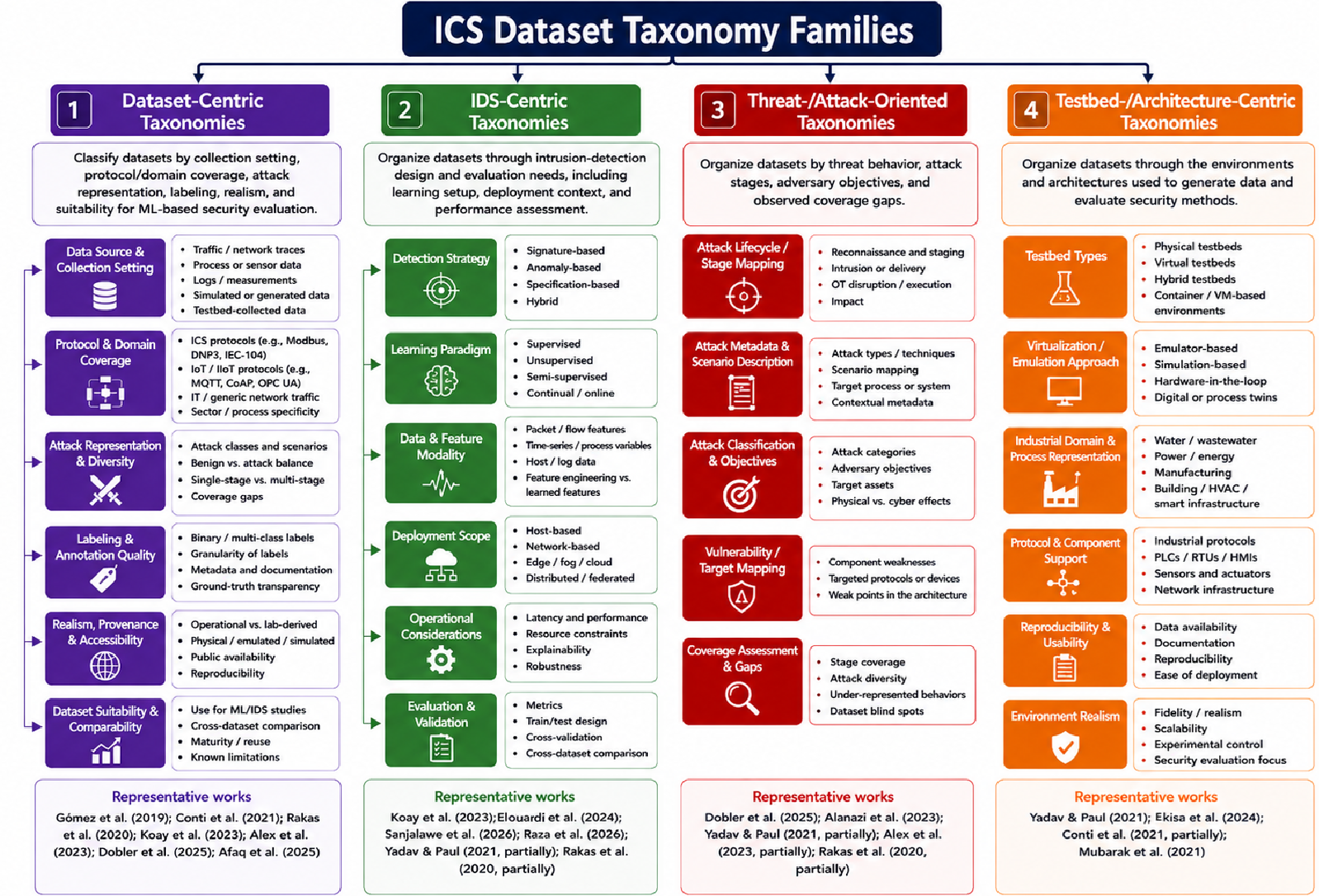}
    \caption{Taxonomy families identified across the reviewed literature.}
    \label{F:Taxonomies}
\end{figure*}

Across the surveyed literature, ICS/OT taxonomies fall broadly into four families: dataset‑centric, IDS‑centric, attack‑centric, and testbed/architecture‑centric schemes. Dataset‑centric taxonomies classify benchmarks by how data are generated and annotated; IDS‑centric schemes organize detection approaches by method and deployment; attack‑centric schemes follow threat stages; and testbed taxonomies describe how cyber‑physical environments are built and instrumented. The complete taxonomy map of the reviewed work is presented in Figure~\ref{F:Taxonomies}.

Dataset-centric taxonomies focus on the characteristics of the datasets themselves. These taxonomies classify benchmarks according to multiple factors like data source, collection methodology, protocol coverage, attack representation, and realism. Reviews of public ICS, IIoT, and smart-grid datasets commonly adopt this perspective to assess dataset suitability for security research and ML applications~\cite{raza2026deeplearning,afaq2025publicdatasets,alex2023iotdatasets,martins2024comparative,alani2023survey,muhammad2024smartgrid,lin2019using,gomez2019generation,conti2021survey,mubarak2021anomaly}. Several studies further distinguish between legacy benchmark datasets and more recent ICS/IIoT collections, emphasizing issues such as class imbalance, feature diversity, and evaluation quality.

IDS-centric taxonomies classify intrusion detection approaches according to their design and operational characteristics. Typical dimensions include detection strategy, learning paradigm, model architecture. Recent surveys differentiate between conventional ML approaches, DL methods, and hybrid or LLM-enhanced architectures~\cite{Sanjalawe2026,koay2023machine,elouardi2024hybridcnnllm,rakas2020review,alanazi2023scada,yadav2021architecture}.

Attack-centric taxonomies organize the literature around adversarial behavior and threat progression. Such schemes classify attacks according to objectives, techniques, or stages of an attack lifecycle, often drawing on kill-chain concepts or vulnerability-based models~\cite{dobler2025systematic,alex2023iotdatasets,alanazi2023scada,yadav2021architecture,rakas2020review}. Their main purpose is to evaluate attack coverage, identify underrepresented threat classes, and assess whether existing datasets and IDS evaluations adequately reflect realistic attack scenarios.

Testbed-centric taxonomies focus on the environments used to generate data and evaluate security solutions. These schemes classify cyber-physical platforms according to their degree of realism, virtualization approach, industrial process representation, protocol support, and reproducibility characteristics~\cite{ekisa2024virtual,conti2021survey,mubarak2021anomaly}. Because datasets and IDS evaluations are strongly influenced by the underlying experimental environment, testbed taxonomies often act as a bridge between dataset-oriented and IDS-oriented classifications.

Overall, as observed in Figure~\ref{F:Taxonomies}, the reviewed studies frequently span multiple taxonomy families. Several dataset-oriented surveys also discuss attack coverage and evaluation criteria, while IDS-focused works often incorporate attack classifications, protocol considerations, and deployment constraints. Likewise, testbed-oriented studies commonly examine the datasets and security evaluations enabled by different experimental environments. The figure also reveals an uneven distribution of research attention. Dataset-centric and IDS-centric perspectives dominate the literature, reflecting the increasing demand for benchmark datasets. In contrast, fewer studies focus primarily on attack modeling or testbed design, even though both strongly influence dataset realism and evaluation validity. Finally, several dimensions recur across multiple taxonomy families. This overlap suggests that datasets, attacks, IDSs, and testbeds should be viewed as closely connected components of the broader ICS/OT cybersecurity ecosystem. 

\section{Unified Taxonomy}
\label{S:unified_taxonomy}

The analysis of prior taxonomies in Section~\ref{S:Taxonomy} reveals a consistent structural problem. Namely, the surveyed works converge on the same core classification concerns, i.e, attack representation, system scope, data provenance, and evaluation quality, but they operationalize these concerns quite differently; they often apply inconsistent terminology and rarely address all of them simultaneously. Specifically, dataset-centric surveys classify by modality, protocol, and attack coverage but treat testbed fidelity as a secondary descriptor~\cite{afaq2025publicdatasets,alex2023iotdatasets,alani2023survey,lin2019using}. On the other hand, IDS-centric surveys organize detection methods but handle datasets as implicit supporting artifacts and not as independently characterized resources~\cite{koay2023machine,mubarak2021anomaly,
rakas2020review,raza2026deeplearning,Sanjalawe2026}. 

Moreover, attack-centric schemes describe threat progression but do not link it systematically to what datasets actually capture~\cite{alanazi2023scada,yadav2021architecture,dobler2025systematic}.
Testbed-centric reviews characterize experimental environments with care, but leave labeling quality and ground-truth provenance largely implicit~\cite{conti2021survey,ekisa2024virtual,gomez2019generation}. Essentially, the consequence is that no existing taxonomy is simultaneously comprehensive, operationally grounded, and consistently applied across the full dataset corpus; think, for example, that two datasets can receive the same informal label ( e.g., multi-attack SCADA network trace), while differing fundamentally in the kill-chain stage coverage, observable evidence layer, testbed fidelity, and label granularity.

The unified taxonomy proposed in this section aims to resolve this fragmentation by grounding each classification dimension in recognized reference frameworks for ICS cybersecurity. Each dimension is derived from a framework that already defines the relevant vocabulary, boundaries, and groupings for the domain. The frameworks and the analytical role each plays in the unified taxonomy are detailed in the following subsection. To the best of our knowledge, this taxonomy enables the first systematic, standard-anchored, like-for-like comparison of the ICS cybersecurity dataset ecosystem.

\subsection{Framework-Derived Taxonomy Dimensions}
\label{SS:taxonomy_dimensions}

The unified taxonomy spans five dimensions, labelled \textbf{D1} through \textbf{D5}, each anchored in one or more of the five reference frameworks listed below. The mapping from framework to dimension is not arbitrary. Specifically, it follows the specific analytical purpose of each standard and the type of dataset attribute it most naturally characterizes. Together, the five dimensions cover the adversary perspective, the system and governance perspective, and the operational consequence perspective, which together constitute the three thematic axes under which the reviewed literature is organized in Section~\ref{SSS:adversary_lifecycle} to~\ref{SSS:operational_impact}.

\begin{enumerate}[label=\textbf{D\arabic*:}]

\item ATT\&CK for ICS~\cite{mitre_ics} defines the tactical goals and associated techniques used to characterize adversary behavior in ICS environments. For corpus-level comparison, we aggregate the ATT\&CK for ICS tactics into three analytical macro-groups: (i) Reconnaissance and Staging, covering preparatory and foothold-building activity such as Initial Access, Execution, Persistence, Discovery, Lateral Movement, Collection, and Command and Control; (ii) OT Disruption, covering Inhibit Response Function and Impair Process Control; and (iii) Impact, retaining the ATT\&CK Impact tactic as the final consequence-oriented group. These macro-groups are introduced in this study for analytical compression and are not native ATT\&CK categories. This dimension records which of the three groups are represented in each dataset.

\item The ICS Cyber Kill Chain~\cite{sans_icskc} divides adversary activity into Stage~1, including cyber intrusion preparation and delivery targeting the IT/enterprise network, and Stage~2, including ICS-specific exploit development, execution, and physical process manipulation or damage. The SANS formulation~\cite{sans_icskc} makes the IT/OT boundary crossing explicit and adds sub-phases for ICS reconnaissance, weaponization, and commissioning. This dimension can be used to classify whether a dataset captures Stage~1 activity, Stage~2 activity, or a continuous cross-stage scenario that crosses the IT/OT boundary.

\item IEC~62443~\cite{iec62443} organizes IACS assets into zones and conduits, namely collections of assets sharing common security requirements and the communication paths connecting them. To make architectural evidence comparable across datasets, we complement this zone-and-conduit perspective with the Purdue Reference Model~\cite{WILLIAMS1994141}, using its hierarchical levels to characterize where observable evidence originates: Level~0 (field devices and process sensors), Level~1 (PLCs and local controllers), Level~2 (supervisory/SCADA and HMI), Level~3 (operations and MES), and Level~3.5/4 (DMZ and enterprise IT). For example, regarding the communication protocol active in each zone, it follows this architecture: Modbus and Profibus at Levels~0--1; DNP3 and S7Comm at Levels~1--2; IEC~60870-5-104 and IEC~61850 at Levels~2--3; standard TCP/IP at Levels~3--4. This dimension can be employed to record the lowest and highest Purdue levels from which a dataset collects observable evidence, and thereby encodes its protocol context and zone coverage.

\item NIST~800-82r3~\cite{nist80082} provides a reference framework for characterizing OT environments, architectures, sectors, and operational contexts. Drawing on this distinction between operational and non-operational OT settings, we define four provenance classes for dataset comparison: operational, physical testbed, emulated/virtualized, and simulation-only. Where appropriate, datasets are also characterized by industrial sector and public accessibility. These classes are analytical categories introduced in this study to capture the degree of operational realism and provenance represented by each dataset.

\item The \textit{Detect} function of the NIST Cybersecurity Framework (CSF)~\cite{nist_csf_2024} requires that anomalies and events be identified (DE.AE), that security monitoring be continuous (DE.CM), and that detection processes be maintained (DE.DP). For a dataset to support credible IDS evaluation, i.e., the applied purpose of the CSF Detect function, its labeling must provide sufficient evidence quality for anomaly identification and evaluation reproducibility. In other words, this dimension can be used to characterize each dataset's label scheme (binary vs.\ multi-class), label granularity (event-level vs.\ run-level), and label provenance (live-executed vs.\ post-hoc injected).

\end{enumerate}

\begin{figure*} [!ht]
    \centering
    \includegraphics[width=1\linewidth]{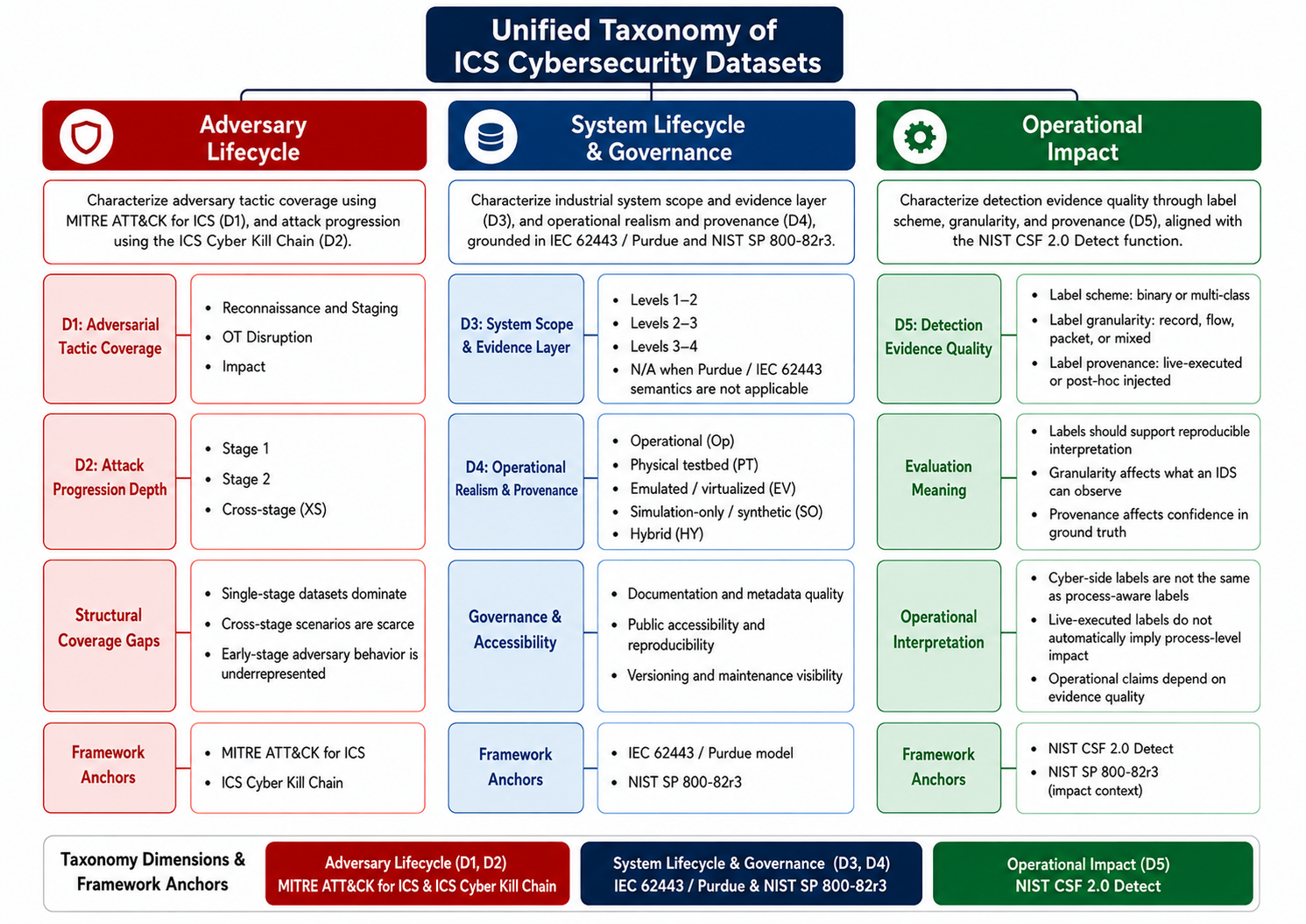}
    \caption{Unified Taxonomy of ICS Cybersecurity Datasets}
    \label{fig:unified_taxonomy}
\end{figure*}

The five dimensions, their respective framework anchors, and the thematic group to which
each primarily belongs to, are summarized in Table~\ref{tab:dimension_framework_map} and illustrated structurally in Figure~\ref{fig:unified_taxonomy}.

\begin{table*}[!htbp]
\centering
\scriptsize
\caption{Taxonomy dimensions, reference framework anchors, and thematic assignments used throughout Section~\ref{S:unified_taxonomy}.}
\label{tab:dimension_framework_map}
\resizebox{\textwidth}{!}{%
\begin{tabular}{@{}llll@{}}
\toprule
\textbf{Dim.} & \textbf{Name} & \textbf{Framework anchor} & \textbf{Theme} \\
\midrule
D1 & Adversarial Tactic Coverage       & MITRE ATT\&CK for ICS~\cite{mitre_ics}          & Adversary \\
D2 & Attack Progression Depth           & ICS Cyber Kill Chain~\cite{sans_icskc} & Adversary \\
D3 & System Scope \& Evidence Layer     & IEC~62443~\cite{iec62443}                         & Governance \\
D4 & Operational Realism \& Provenance  & NIST SP~800-82r3~\cite{nist80082}                 & Governance \\
D5 & Detection Evidence Quality         & NIST CSF Detect~\cite{nist_csf_2024}                   & Impact \\
\bottomrule
\end{tabular}}
\end{table*}

\subsection{Adversary Lifecycle}
\label{SSS:adversary_lifecycle}

Dimensions \textbf{D1} and \textbf{D2} jointly characterize the adversary perspective on ICS datasets: \textbf{D1} records which ATT\&CK for ICS tactic macro-stages a dataset provides evidence for, while \textbf{D2} records whether that evidence spans Stage~1, Stage~2, or a cross-stage scenario according to the ICS Kill Chain. Works in this thematic group therefore focus on attack structure, lifecycle coverage, and adversary-behavior classification.

The authors in~\cite{dobler2025systematic} examine 97 attack types across 32 open-access IT/IoT/OT/ICS datasets and map them to Cyber Kill Chain stages, showing that later-stage activity is considerably better represented than earlier intrusion preparation and progression. The work in~\cite{martins2024comparative} similarly finds that realistic multi-phase attacks crossing the IT/OT boundary are absent from its surveyed corpus and that Stage~2 coverage favors denial-of-availability events over process integrity violations and safety-system inhibition. The authors in~\cite{afaq2025publicdatasets} confirm the corresponding \textbf{D1} imbalance, with ATT\&CK for ICS coverage concentrated on Impact and Inhibit Response Function while Initial Access, Execution, Persistence, and Lateral Movement remain scarcely represented. Together, these findings indicate that IDS evaluation based on individual public datasets is typically restricted to a narrow portion of the ATT\&CK for ICS tactic space.

The work in~\cite{alanazi2023scada} further distinguishes disruption-oriented attacks, including DoS, replay, and command injection, from deception and integrity-violation attacks such as FDI and MitM, showing that disruption remains overrepresented even within Stage~2. In the smart-grid domain,~\cite{alani2023survey} likewise reports the underrepresentation of FDI and MitM attacks, while~\cite{yadav2021architecture} separates cyber-layer from cyber-physical attacks in a manner consistent with the Stage~1/Stage~2 distinction of \textbf{D2}. The authors in~\cite{alex2023iotdatasets} identify a similar lifecycle-completeness problem in IoT and IIoT datasets, where large numbers of labeled instances remain concentrated within a narrow behavioral envelope. Finally, the work in~\cite{koay2023machine} links these \textbf{D1} and \textbf{D2} gaps directly to ML evaluation validity, arguing that narrow kill-chain coverage limits generalization to multi-stage threat scenarios. Collectively, these works show that the central adversary-lifecycle limitation is not the absence of datasets, but the lack of tactic diversity and multi-stage progression across the corpus.

\begin{mdframed}[backgroundcolor=gray!10, linecolor=black, linewidth=0.5pt, roundcorner=5pt]
\textit{\textbf{Argument \#1}: Public ICS datasets systematically overrepresent late-stage disruption while leaving early-stage intrusion, persistence, and cross-boundary progression largely uncovered, yet this imbalance is invisible behind generic attack-type labels that conflate tactic diversity with instance count. \textbf{D1} and \textbf{D2} are the minimum necessary dimensions to make this gap measurable and comparable across datasets.}
\end{mdframed}

\subsection{System Lifecycle and Governance}
\label{SSS:system_lifecycle}

Dimensions \textbf{D3} and \textbf{D4} jointly characterize the governance perspective on ICS datasets. \textbf{D3} records the observable evidence layer using the IEC~62443 Zones-and-Conduits model and Purdue Reference Model levels, while \textbf{D4} records operational realism and provenance using NIST~800-82r3 system topology and sector categories. Works in this thematic group therefore focus on system architecture, testbed fidelity, protocol context, dataset accessibility, and lifecycle governance.

The work in~\cite{yadav2021architecture} provides an architecturally explicit analysis by mapping ICS components to Purdue levels and showing that public datasets are concentrated at Levels~2--3, while Levels~0--1 remain sparsely represented, exposing a clear \textbf{D3} gap in field-level evidence. The authors in~\cite{conti2021survey} examine testbeds and datasets through system composition, component roles, communication topology, and fidelity, and recommend stronger process-level evidence in addition to network-layer traffic. From the \textbf{D4} perspective, the work in~\cite{ekisa2024virtual} evaluates 21 virtual ICS testbeds across virtualization technology, protocol coverage, behavioral realism, and reproducibility, showing that software-based emulation may differ from physical systems in timing, protocol behavior, and process dynamics. Similarly, the authors in~\cite{gomez2019generation} make dataset provenance explicit through a four-stage methodology covering attack selection, deployment, traffic capture, and feature computation, demonstrated through the Electra dataset. The work in~\cite{rakas2020review} further evaluates SCADA IDS testbeds according to network segmentation, traffic authenticity, protocol correctness, and related criteria, finding that most satisfy only a subset of these requirements.

Protocol and sector context provide an additional \textbf{D3} concern. The authors in~\cite{muhammad2024smartgrid} show in the smart-grid domain that dataset characterization must retain protocol and sector information, since operational requirements differ across environments and communication technologies. Similarly, the work in~\cite{alani2023survey} finds that protocol diversity in public smart-grid datasets is substantially narrower than in real operational environments, which limits generalizability. At the governance level,~\cite{afaq2025publicdatasets} examines documentation quality, licensing, accessibility, and maintenance status across public ICS and OT datasets, and shows that incomplete or poorly maintained resources weaken their usefulness for evaluation and continuous monitoring. Finally, the authors in~\cite{lin2019using} highlight how repeated reuse of stale or poorly documented benchmarks can perpetuate datasets that no longer reflect current operational environments, protocol stacks, or threat models.

\begin{mdframed}[backgroundcolor=gray!10, linecolor=black, linewidth=0.5pt, roundcorner=5pt]
\textit{\textbf{Argument \#2}: Public ICS datasets are not only adversarially narrow but also architecturally shallow and operationally weak as they mainly capture supervisory-layer traffic while field-level evidence remains sparse, are often generated in emulated environments with limited fidelity reporting, and frequently omit the protocol, sector, and lifecycle information needed for governance-aware evaluation. Therefore, \textbf{D3} and \textbf{D4} are essential for distinguishing datasets that are merely available from those that are architecturally representative and operationally credible.}
\end{mdframed}

\subsection{Operational Impact}
\label{SSS:operational_impact}

Dimension \textbf{D5} characterizes the quality and completeness of dataset ground truth, anchored in the NIST CSF Detect function~\cite{nist_csf_2024} as operationalized in NIST~800-82r3. The Detect function requires identification of anomalies and events (DE.AE), continuous monitoring (DE.CM), and maintained detection processes (DE.DP), making evidence quality, label granularity, and reproducibility central to credible IDS evaluation. The work in~\cite{lin2019using} emphasizes synchronized physical-process measurements and network traffic so that detection can be assessed against both cyber activity and physical consequences. Similarly,~\cite{mubarak2021anomaly} evaluates ML classifiers on SCADA and ICS process-variable records, showing the distinct value of physical-process evidence, while the authors in~\cite{Sanjalawe2026} identify the lack of high-fidelity cyber-physical datasets as a barrier to continuous monitoring in realistic IIoT environments.

Label granularity constitutes another major \textbf{D5} concern. The work in~\cite{dobler2025systematic} identifies coarse labeling and broad attack windows as limitations for precise anomaly interpretation, while~\cite{afaq2025publicdatasets} highlights labeling quality as a recurring dataset weakness. Similarly,~\cite{martins2024comparative} reports inconsistent labeling and limited granularity across OT and ICS datasets. The authors in~\cite{koay2023machine} connect this problem directly to ML evaluation, arguing that aggregate performance measures on coarse labels provide limited information about detection of specific attack techniques. The work in~\cite{elouardi2024hybridcnnllm} further shows that evaluation outcomes vary across datasets with different labeling schemes, including binary versus multi-class labels and different ground-truth characteristics. Together, these studies show that dataset size alone cannot compensate for labels that fail to identify individual events, attack timing, or physical-process consequences.

From the dataset-generation perspective, the authors in~\cite{gomez2019generation} address \textbf{D5} by constructing controlled attack scenarios with known provenance, enabling more reproducible ground truth for physical-impact events. The work in~\cite{ekisa2024virtual} complements this from the testbed perspective, emphasizing that meaningful physical-impact labels require sufficiently faithful reproduction of process dynamics rather than network-layer emulation alone. In the energy domain,~\cite{muhammad2024smartgrid} examines smart-grid attacks and datasets in relation to operational and protocol-specific consequences, while the authors in~\cite{raza2026deeplearning} discuss IIoT attack scenarios and the need for datasets that better reflect realistic industrial conditions and evolving threats. Finally, the survey in~\cite{conti2021survey} highlights the relative maturity of energy-sector testbeds and datasets compared with other industrial sectors, exposing an uneven basis for characterizing physical consequences across the broader ICS corpus.

\begin{mdframed}[backgroundcolor=gray!10, linecolor=black, linewidth=0.5pt, roundcorner=5pt]
\textit{\textbf{Argument \#3}: Credible assessment of operational impact requires synchronized cyber-physical evidence and event-level ground truth, but the public ICS corpus is dominated by run-level labels, limited process visibility, and emulated dynamics that blur the physical consequences of attacks. Hence, \textbf{D5} is essential for separating datasets that support only attack detection from those that support reproducible measurement of operational impact.}
\end{mdframed}

\subsection{Synthesis and Taxonomic Value}
\label{SSS:synthesis}

Taken together, the three arguments made in subsections~\ref{SSS:adversary_lifecycle} to~\ref{SSS:operational_impact} capture the essence of our analysis. Namely, \textbf{Argument \#1} shows that public ICS datasets are adversarially narrow, \textbf{Argument \#2} shows that they are architecturally shallow and operationally under-specified, and \textbf{Argument \#3} shows that they provide insufficient ground truth for measuring operational impact. In this context, we claim that the proposed taxonomy is stronger than prior schemes because it separates these weaknesses into standard-anchored dimensions with clear analytical roles, while being grounded in well-established ICS security standards and guidelines. 

Specifically, \textbf{D1} and \textbf{D2} make adversary coverage measurable, \textbf{D3} and \textbf{D4} make system scope and provenance explicit, and \textbf{D5} makes evidence quality visible. This framework provides a unified basis for comparing heterogeneous datasets while exposing the structural gaps that limit realism, reproducibility, and generalization. In this sense, the taxonomy in Figure~\ref{fig:unified_taxonomy} does not replace earlier schemes, but reconciles them into a single attack-, system-, and impact-aware structure that is better suited to dataset comparison, construction, and evaluation.

\begin{table*}[!htbp]
\small
\centering
\caption{Unified taxonomy of datasets identified across the included literature. D1: R\&S = Reconnaissance and Staging, OTD = OT Disruption, Imp = Impact.
D2: S1 = Stage~1, S2 = Stage~2, XS = cross-stage or multi-stage scenario.
D3: Purdue evidence layer; N/A indicates that Purdue-level semantics are not applicable.
D4: Op = operational, PT = physical testbed, EV = emulated/virtualized, SO = simulation-only/synthetic, HY = hybrid.
D5: Bin = binary labels, MC = multi-class labels, LE = live-executed labels, PHI = post-hoc injected labels.}
\label{tab:dataset_taxonomy_full}
\resizebox{\textwidth}{!}{%
\begin{tabular}{@{}l l c c c c c c c@{}}
\toprule
\multirow{2}{*}{\textbf{Dataset}}
  & \multirow{2}{*}{\textbf{Work}}
  & \textbf{D1} & \textbf{D2} & \textbf{D3} & \textbf{D4}
  & \multicolumn{3}{c}{\textbf{D5}} \\
\cmidrule(lr){7-9}
  &  & \makecell[c]{\scriptsize Tactic\\\scriptsize coverage}
  & \makecell[c]{\scriptsize Progression\\\scriptsize depth}
  & \makecell[c]{\scriptsize Evidence\\\scriptsize layer}
  & \makecell[c]{\scriptsize Provenance}
  & \makecell[c]{\scriptsize Scheme}
  & \makecell[c]{\scriptsize Granularity}
  & \makecell[c]{\scriptsize Label\\\scriptsize provenance} \\
\midrule

\multicolumn{9}{@{}c}{\textbf{ICS / OT datasets}}\\
\midrule

D5 Energy M.S.D. (EMS logs)
  & \cite{conti2021survey}
  & N/A & N/A & 3--4 & Op
  & N/A & Record & LE \\

Power System (Morris)
  & \cite{conti2021survey}, \cite{dobler2025systematic}, \cite{martins2024comparative}
  & OTD, Imp & S2 & 2--3 & PT
  & Bin/MC & Record & LE \\

Gas Pipeline (Morris)
  & \cite{conti2021survey}, \cite{dobler2025systematic}, \cite{koay2023machine}
  & R\&S, OTD, Imp & S2 & 1--2 & PT
  & MC & Record & LE \\

Water Storage Tank
  & \cite{conti2021survey}
  & R\&S, OTD & S2 & 1--2 & PT
  & MC & Record & LE \\

New Gas Pipeline
  & \cite{conti2021survey}, \cite{dobler2025systematic}
  & OTD & S2 & 1--2 & PT
  & MC & Record & LE \\

BATADAL
  & \cite{lin2019using}, \cite{conti2021survey}, \cite{dobler2025systematic}, \cite{martins2024comparative}, \cite{koay2023machine}
  & R\&S, OTD, Imp & XS & 2--3 & SO
  & MC & Record & LE \\

SWaT
  & \cite{lin2019using}, \cite{conti2021survey}, \cite{dobler2025systematic}, \cite{martins2024comparative}, \cite{raza2026deeplearning}, \cite{Sanjalawe2026}, \cite{afaq2025publicdatasets}, \cite{alanazi2023scada}, \cite{alex2023iotdatasets}, \cite{koay2023machine}, \cite{rakas2020review}, \cite{mubarak2022industrial}, \cite{yadav2021architecture}
  & R\&S, OTD, Imp & XS & 1--2 & PT
  & Bin/MC & Record & LE \\

WADI
  & \cite{lin2019using}, \cite{conti2021survey}, \cite{dobler2025systematic}, \cite{martins2024comparative}, \cite{raza2026deeplearning}, \cite{Sanjalawe2026}, \cite{afaq2025publicdatasets}, \cite{alanazi2023scada}, \cite{koay2023machine}
  & OTD, Imp & S2 & 1--2 & PT
  & Bin/MC & Record & LE \\

EPIC
  & \cite{lin2019using}, \cite{conti2021survey}, \cite{dobler2025systematic}, \cite{martins2024comparative}, \cite{afaq2025publicdatasets}, \cite{mubarak2022industrial}
  & OTD, Imp & S2 & 2--3 & PT
  & MC & Record & LE \\

HAI
  & \cite{conti2021survey}, \cite{martins2024comparative}, \cite{Sanjalawe2026}, \cite{afaq2025publicdatasets}, \cite{koay2023machine}, \cite{mubarak2022industrial}
  & OTD, Imp & S2 & 1--2 & PT
  & Bin/MC & Record & LE \\

Electra
  & \cite{gomez2019generation}, \cite{conti2021survey}, \cite{dobler2025systematic}, \cite{martins2024comparative}, \cite{koay2023machine}
  & R\&S, OTD & S2 & 1--2 & PT
  & Bin/MC & Packet/Flow & LE \\

QUT\_DNP3
  & \cite{conti2021survey}, \cite{dobler2025systematic}, \cite{muhammad2024smartgrid}
  & R\&S, OTD & S2 & 1--2 & PT
  & MC & Packet/Flow & LE \\

QUT\_S7Comm
  & \cite{conti2021survey}, \cite{dobler2025systematic}
  & OTD & S2 & 1--2 & PT
  & MC & Packet/Log & LE \\

Modbus SCADA \#1
  & \cite{conti2021survey}, \cite{dobler2025systematic}
  & OTD & S2 & 1--2 & PT
  & MC & Packet/Flow & LE \\

Modbus/TCP Dataset
  & \cite{alex2023iotdatasets}, \cite{dobler2025systematic}
  & OTD & S2 & 1--2 & PT
  & Bin/MC & Packet/Record & LE \\

Lemay SCADA
  & \cite{conti2021survey}
  & R\&S, OTD & S2 & 2--3 & EV
  & MC & Packet/Flow & PHI \\

Lemay Covert
  & \cite{conti2021survey}
  & OTD & S2 & 2--3 & EV
  & Bin & Packet/Flow & PHI \\

4SICS
  & \cite{conti2021survey}, \cite{dobler2025systematic}
  & R\&S, OTD & S1 & 2--3 & PT
  & N/A & Packet & LE \\

S4x15 ICS
  & \cite{conti2021survey}
  & R\&S, OTD & S1 & 2--3 & PT
  & N/A & Packet & LE \\

CyberCity
  & \cite{conti2021survey}, \cite{dobler2025systematic}
  & R\&S, OTD & XS & 2--3 & PT
  & N/A & Packet & LE \\

HVAC Traces
  & \cite{conti2021survey}
  & N/A & N/A & 3--4 & Op
  & N/A & Packet & LE \\

Tennessee Eastman Process
  & \cite{koay2023machine}, \cite{dobler2025systematic}
  & Imp & S2 & 1--2 & SO
  & MC & Record & LE \\

TLIGHT
  & \cite{koay2023machine}
  & OTD, Imp & S2 & 1--2 & PT
  & MC & Record & LE \\

DNP3 Intrusion Dataset
  & \cite{alani2023survey}, \cite{muhammad2024smartgrid}, \cite{dobler2025systematic}
  & R\&S, OTD & S2 & 1--2 & PT
  & MC & Flow & LE \\

IEC 60870-5-104 Intrusion Dataset
  & \cite{alani2023survey}, \cite{muhammad2024smartgrid}, \cite{dobler2025systematic}
  & R\&S, OTD & S2 & 1--2 & PT
  & MC & Flow & LE \\

OPCUA Dataset
  & \cite{alex2023iotdatasets}, \cite{dobler2025systematic}
  & OTD & S2 & 1--2 & PT
  & Bin/MC & Record & LE \\

DoS and MitM Dataset PLC
  & \cite{alex2023iotdatasets}
  & OTD & S2 & 1--2 & PT
  & Bin/MC & Flow & LE \\

KNX Datasets
  & \cite{alex2023iotdatasets}
  & OTD & S2 & 2--3 & HY
  & N/A & Packet/Flow & LE \\

Mubarak ICS SCADA Cyber Kit 
 & \cite{mubarak2021anomaly}, \cite{dobler2025systematic} 
 & R\&S, OTD & S2 & 1--2 & PT 
 & Bin/MC & Packet/Record & LE \\

\midrule
\multicolumn{9}{@{}c}{\textbf{IIoT / industrial network-security datasets}}\\
\midrule

WUSTL-IIoT-2018
  & \cite{conti2021survey}, \cite{dobler2025systematic}, \cite{martins2024comparative}, \cite{alex2023iotdatasets}, \cite{Sanjalawe2026}
  & OTD & S2 & 2--3 & PT
  & Bin & Flow & LE \\

WUSTL-IIoT-2021
  & \cite{martins2024comparative}, \cite{alex2023iotdatasets}, \cite{Sanjalawe2026}
  & R\&S, OTD & S2 & 2--3 & PT
  & Bin/MC & Flow & LE \\

X-IIoTID
  & \cite{martins2024comparative}, \cite{alex2023iotdatasets}, \cite{elouardi2024hybridcnnllm}, \cite{Sanjalawe2026}, \cite{dobler2025systematic}
  & R\&S, OTD, Imp & XS & 2--3 & PT
  & MC & Flow & LE \\

ICS-Flow
  & \cite{martins2024comparative}, \cite{muhammad2024smartgrid}
  & R\&S, OTD & S2 & 2--3 & PT
  & MC & Flow & LE \\

TON\_IoT
  & \cite{raza2026deeplearning}, \cite{Sanjalawe2026}, \cite{afaq2025publicdatasets}, \cite{elouardi2024hybridcnnllm}, \cite{alex2023iotdatasets}
  & R\&S, OTD, Imp & XS & 2--3 & HY
  & Bin/MC & Record/Flow & LE \\

Edge-IIoTset
  & \cite{Sanjalawe2026}, \cite{afaq2025publicdatasets}, \cite{elouardi2024hybridcnnllm}, \cite{alex2023iotdatasets}, \cite{martins2024comparative}
  & R\&S, OTD, Imp & S2 & 2--3 & HY
  & MC & Flow & LE \\

CIC APT IIoT 2024
  & \cite{afaq2025publicdatasets}
  & R\&S, OTD, Imp & XS & 2--3 & EV/PT
  & MC & Flow & LE \\

Gotham Testbed Dataset
  & \cite{afaq2025publicdatasets}
  & R\&S, OTD & S2 & 2--3 & EV
  & MC & Flow & LE \\

Blaq\_0 Hackathon
  & \cite{afaq2025publicdatasets}
  & R\&S, OTD & S2 & 2--3 & PT
  & MC & Packet & LE \\

HDGM
  & \cite{dobler2025systematic}
  & R\&S, OTD & S2 & 2--3 & EV
  & MC & Flow & LE \\

Building Power Consumption Dataset
  & \cite{afaq2025publicdatasets}
  & N/A & N/A & 3--4 & Op
  & N/A & Record & LE \\

LBNL-ETA/Brick Dataset
  & \cite{afaq2025publicdatasets}
  & N/A & N/A & 3--4 & Op
  & N/A & Record & LE \\

\midrule
\multicolumn{9}{@{}c}{\textbf{IoT security datasets}}\\
\midrule

IoT-23
  & \cite{afaq2025publicdatasets}, \cite{elouardi2024hybridcnnllm}, \cite{alex2023iotdatasets}
  & R\&S, OTD & S1 & N/A & Op/EV
  & MC & Scenario/Packet & LE \\

N-BaIoT
  & \cite{afaq2025publicdatasets}, \cite{alex2023iotdatasets}, \cite{elouardi2024hybridcnnllm}, \cite{Sanjalawe2026}
  & OTD & S2 & N/A & PT
  & MC & Record & LE \\

BoT-IoT
  & \cite{alex2023iotdatasets}, \cite{elouardi2024hybridcnnllm}, \cite{Sanjalawe2026}
  & R\&S, OTD, Imp & S2 & N/A & EV
  & MC & Flow & LE \\

IoTID20
  & \cite{alex2023iotdatasets}
  & R\&S, OTD & S2 & N/A & PT
  & MC & Packet/Flow & LE \\

IoTID
  & \cite{alex2023iotdatasets}
  & R\&S, OTD & S2 & N/A & PT
  & MC & Packet & LE \\

MQTT-IoT-IDS2020
  & \cite{alex2023iotdatasets}
  & R\&S, OTD & S2 & N/A & PT/EV
  & MC & Flow & LE \\

MQTTset 
 & \cite{alex2023iotdatasets} 
 & R\&S, OTD & S2 & N/A & EV 
 & MC & Flow & LE \\ 

BoTIoTT On IoT 
 & \cite{alex2023iotdatasets} 
 & R\&S, OTD, Imp & S2 & N/A & SO/HY 
 & MC & Flow & PHI \\ 

BoTNeT IoT-L01 
 & \cite{alex2023iotdatasets} 
 & R\&S, OTD & S2 & N/A & SO 
 & MC & Flow & PHI \\

Network Dataset MQTT IoT
  & \cite{alex2023iotdatasets}
  & R\&S, OTD & S2 & N/A & EV/PT
  & MC & Record & LE \\

Kitsune
  & \cite{alex2023iotdatasets}, \cite{Sanjalawe2026}
  & OTD & S1 & N/A & PT
  & Bin & Record & LE \\

IOT-BDA
  & \cite{alex2023iotdatasets}
  & R\&S, OTD & S2 & N/A & Op/EV
  & MC & Packet/JSON & LE \\

MalwareSpecSys
  & \cite{alex2023iotdatasets}
  & OTD & S1 & N/A & PT
  & MC & Syscall/Log & LE \\

NSS Mirai
  & \cite{alex2023iotdatasets}
  & OTD, Imp & S1 & N/A & EV
  & MC & Record & LE \\

IoTDOS and DDoS
  & \cite{alex2023iotdatasets}
  & OTD, Imp & S1 & N/A & SO
  & Bin/MC & Image/Record & PHI \\

TCP FIN Flood and ZBAssocFlood
  & \cite{alex2023iotdatasets}
  & OTD & S1 & N/A & PT
  & Bin/MC & Flow & LE \\

Malicious Network Traffic PCAPs and Binary Images
  & \cite{alex2023iotdatasets}
  & OTD & S1 & N/A & SO/EV
  & Bin/MC & Packet/Image & PHI \\

CICIoT2023
  & \cite{elouardi2024hybridcnnllm}, \cite{Sanjalawe2026}, \cite{afaq2025publicdatasets}
  & R\&S, OTD & S2 & N/A & EV
  & MC & Flow & LE \\

IoT Honeypot Dataset
  & \cite{afaq2025publicdatasets}
  & R\&S, OTD & S1 & N/A & Op
  & Bin/MC & Packet/JSON & LE \\

VPN-Forwarded IoT Honeypot
  & \cite{alex2023iotdatasets}
  & R\&S, OTD & S1 & N/A & Op
  & Bin/MC & Packet/JSON & LE \\

IoTKeeper
  & \cite{alex2023iotdatasets}
  & R\&S, OTD & S1 & N/A & Op
  & Bin/MC & Packet & LE \\

IoTSentinel
  & \cite{alex2023iotdatasets}
  & R\&S & S1 & N/A & Op
  & Bin & Packet/Record & LE \\

DIoT
  & \cite{alex2023iotdatasets}
  & OTD & S1 & N/A & Op/PT
  & Bin & Packet & LE \\

DS2OS Traffic Traces
  & \cite{alex2023iotdatasets}
  & N/A & N/A & N/A & SO
  & Bin/MC & Record & LE \\

MedBIoT
  & \cite{alex2023iotdatasets}
  & OTD & S1 & N/A & PT/EV
  & MC & Packet & LE \\

Urban\_IoT\_DDoS
  & \cite{alex2023iotdatasets}, \cite{elouardi2024hybridcnnllm}
  & OTD, Imp & S1 & N/A & SO
  & Bin/MC & Record & LE \\

CCD-INID-V1
  & \cite{alex2023iotdatasets}
  & R\&S, OTD & S2 & N/A & PT/EV
  & MC & Flow & LE \\

IoT-deNAT
  & \cite{alex2023iotdatasets}
  & R\&S, OTD & S1 & N/A & Op
  & Bin/MC & Record & LE \\

UNSW-IoT
  & \cite{alex2023iotdatasets}
  & R\&S, OTD & S2 & N/A & EV
  & MC & Packet/Flow & LE \\

IEEE TMC 2018 UNSW
  & \cite{alex2023iotdatasets}
  & N/A & N/A & N/A & Op
  & N/A & Packet & LE \\

SoK Dataset
  & \cite{alex2023iotdatasets}
  & N/A & N/A & N/A & Op
  & N/A & Packet & LE \\

IoTInspector
  & \cite{alex2023iotdatasets}
  & N/A & N/A & N/A & Op
  & N/A & Packet & LE \\

IoTFinder
  & \cite{alex2023iotdatasets}
  & R\&S & S1 & N/A & Op
  & Bin & Packet & LE \\

\midrule
\multicolumn{9}{@{}c}{\textbf{General IDS baseline datasets}}\\
\midrule

KDDCup99
  & \cite{raza2026deeplearning}, \cite{Sanjalawe2026}
  & OTD & S2 & N/A & SO
  & MC & Record & PHI \\

NSL-KDD
  & \cite{raza2026deeplearning}, \cite{Sanjalawe2026}, \cite{alani2023survey}, \cite{muhammad2024smartgrid}
  & OTD & S2 & N/A & SO
  & MC & Record & PHI \\

UNSW-NB15
  & \cite{raza2026deeplearning}, \cite{Sanjalawe2026}, \cite{alani2023survey}, \cite{muhammad2024smartgrid}
  & R\&S, OTD & S2 & N/A & EV
  & MC & Flow & LE \\

ISCX-2012
  & \cite{alani2023survey}, \cite{muhammad2024smartgrid}
  & R\&S, OTD & S2 & N/A & EV
  & MC & Flow & LE \\

CICIDS2017
  & \cite{afaq2025publicdatasets}, \cite{Sanjalawe2026}, \cite{raza2026deeplearning}
  & R\&S, OTD & S2 & N/A & EV
  & MC & Flow & LE \\

CSE-CIC-IDS2018
  & \cite{elouardi2024hybridcnnllm}, \cite{Sanjalawe2026}
  & R\&S, OTD & S2 & N/A & EV
  & MC & Flow & LE \\

CICDDoS2019
  & \cite{alex2023iotdatasets}
  & OTD, Imp & S1 & N/A & EV
  & MC & Flow & LE \\

DARPA IDS 
 & \cite{alanazi2023scada} 
 & N/A & N/A & N/A & SO 
 & MC & Packet/Connection & LE \\

NF-UQ-NIDS
  & \cite{alex2023iotdatasets}
  & R\&S, OTD, Imp & XS & N/A & SO
  & Bin/MC & Flow & PHI \\

\bottomrule
\end{tabular}%
}
\end{table*}


\section{Comparative Analysis of Datasets}
\label{S:comparative_dataset_analysis}

The preceding sections established the need for a standard-anchored taxonomy capable of comparing heterogeneous ICS, IIoT, IoT, and IDS datasets on common analytical grounds. This section applies that taxonomy to the recurring dataset corpus to move from framework construction to corpus-level diagnosis. Table~\ref{tab:dataset_taxonomy_full} applies the unified taxonomy to the dataset corpus extracted from the 18 studies, characterizing each dataset across the five dimensions (\textbf{D1}--\textbf{D5}) introduced in Section~\ref{S:unified_taxonomy}. The table is not intended as a new dataset catalogue. Instead, it provides a standard-anchored characterization of the recurring datasets and evaluation evidence discussed across the surveyed literature. Its purpose is to expose corpus-level structure, namely, patterns that remain difficult to observe when datasets are described in isolation or through inconsistent attribute sets.

\begin{figure*}[!ht]
  \centering
  \includegraphics[width=\linewidth]{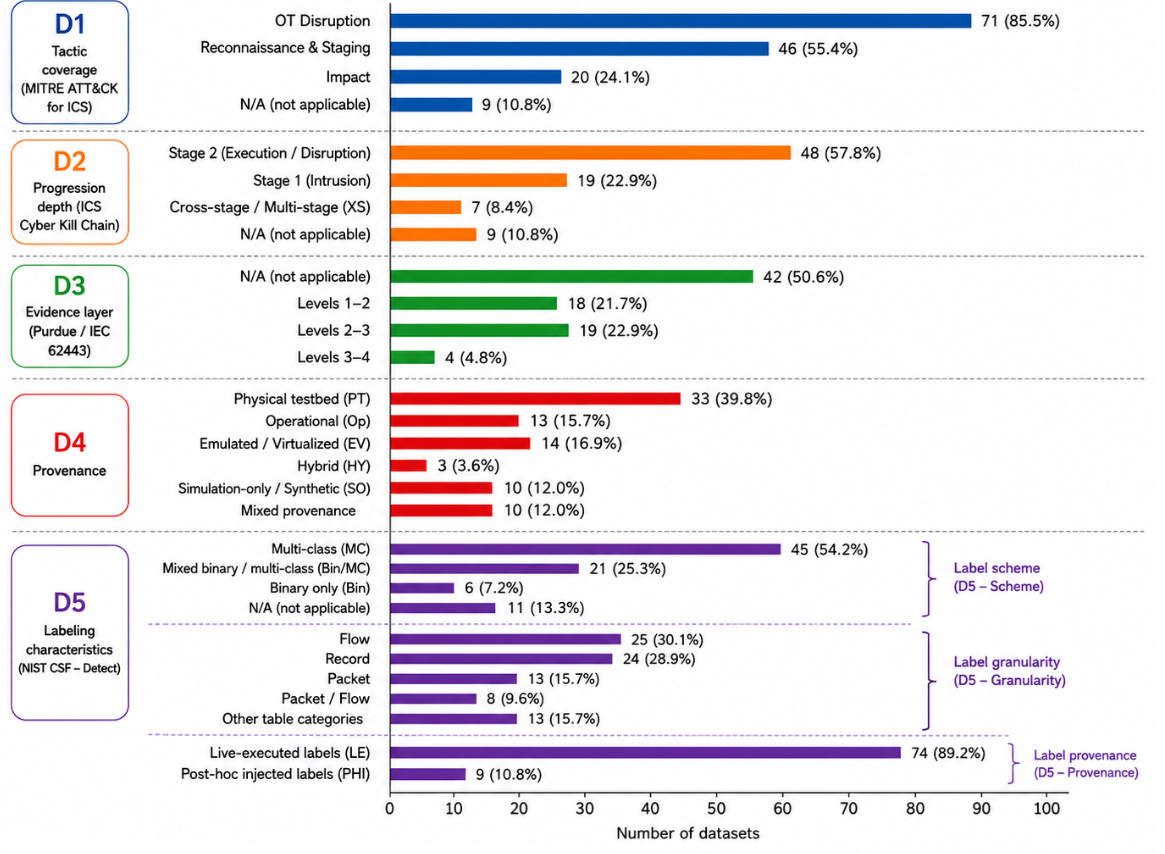}
  \caption{Distribution of key taxonomy characteristics across 83 datasets. \textbf{D1} is multi-label; therefore, its percentages do not sum to 100\%.}
  \label{fig:taxonomy_distribution}
\end{figure*}

The resulting corpus contains 83 dataset rows, and read as a whole, it is structurally uneven. It is dominated by disruption-oriented, Stage~2-centric, and laboratory-derived evidence, while explicit Purdue-layer coverage, operational provenance, and process-aware ground truth remain comparatively scarce. Figure~\ref{fig:taxonomy_distribution} summarizes this distribution across \textbf{D1--D5}. The table provides the dataset-level classification; the figure exposes the aggregate skew produced by those classifications. In this respect, the counts reported below should be interpreted descriptively. They do not test statistical significance, but they reveal systematic coverage patterns in what the public dataset ecosystem makes observable.

\subsection{Adversarial Coverage and Progression}
\label{S:Comparative:D1D2}

The clearest imbalance appears in the adversary-facing dimensions \textbf{D1} and \textbf{D2}. In \textbf{D1}, OT~Disruption is represented in 85.5\% of the datasets, making it by far the dominant adversarial macro-stage. Reconnaissance and Staging appears in 46 datasets, whereas Impact appears in only 20. This distribution indicates that public benchmarks predominantly capture attacks that interfere with industrial control operations, but less frequently represent either the preparatory behaviours that precede disruption or the resulting operational and physical consequences, such as service interruption, unsafe process conditions, production loss, or equipment degradation~\cite{dobler2025systematic,afaq2025publicdatasets,alanazi2023scada}.

A similar pattern appears in \textbf{D2}. Stage~2 scenarios account for 57.8\% of the datasets, whereas only 8.4\% are coded as cross-stage. Thus, even when datasets contain multiple attack types, they usually represent isolated disruption episodes and not an end-to-end intrusion progression across the IT/OT boundary. This constitutes a progression-compression imbalance: IDS models can be evaluated extensively on late-stage disturbances, yet remain weakly tested against intrusion buildup, staging, lateral movement, and boundary-crossing behaviour~\cite{martins2024comparative,dobler2025systematic,koay2023machine}.

The implication is not merely that early-stage attacks are underrepresented. On the contrary, the corpus provides limited evidence for the relationship between early adversary activity and later process disruption. A model trained primarily on Stage~2 windows may learn the signatures of process disturbance without learning the behaviours that enable such disturbance. Consequently, \textbf{D1} and \textbf{D2} expose a central limitation of the public dataset ecosystem: attack volume is not equivalent to adversary lifecycle coverage.

\subsection{Architectural Evidence and Purdue-Layer Coverage}
\label{S:Comparative:D3}

The sharpest structural divide appears in \textbf{D3}. Overall, 50.6\% of the datasets are marked as \textit{N/A}, meaning that Purdue/IEC~62443 semantics do not apply to them. This is not, by itself, a defect of those datasets, since many IoT and general IDS benchmarks were never designed to model industrial zones, conduits, PLCs, field devices, supervisory control layers, or process-control interactions. Nevertheless, their \textbf{D3}=\textit{N/A} status has a direct interpretive consequence: they cannot be treated as substitutes for process-aware ICS evidence.

Among the 41~datasets where \textbf{D3} is applicable, architectural visibility is concentrated mainly at Levels~1--2 and 2--3. Specifically, 18~datasets (21.7\% of the full corpus) are coded at Levels~1--2, and 19~datasets (22.9\%) at Levels~2--3, while only four datasets reach Levels~3--4. Explicit Level~0 evidence in the Purdue hierarchy is absent from Table~\ref{tab:dataset_taxonomy_full}. This is a consequential gap because Level~0 is where the physical process is sensed and actuated. Prior ICS dataset and testbed surveys similarly emphasize that credible industrial security evaluation depends on system architecture, component roles, communication topology, and the relation between network evidence and physical process behaviour~\cite{conti2021survey,yadav2021architecture,rakas2020review}.

This distribution clarifies why broad dataset counts can be misleading. The corpus appears large, but the subset that simultaneously supports industrial architecture, process interpretation, and attack progression is much smaller. Without field-device and process-interaction evidence, datasets can support network-level or controller-level detection, but they provide limited support for evaluating physical-impact detection. Therefore, comparative studies that group process-aware ICS datasets, industrial-style IIoT datasets, generic IoT datasets, and general IDS baselines under a common ``security dataset'' umbrella risk overstating the practical diversity of the evidence base~\cite{alex2023iotdatasets,afaq2025publicdatasets,dobler2025systematic}.

\subsection{Operational Realism and Data Provenance}
\label{S:Comparative:D4}

Dimension \textbf{D4} shows that the corpus remains predominantly laboratory-derived. Physical testbeds form the largest provenance class, accounting for 39.8\% of the datasets. Operational datasets account for 15.7\%, emulated or virtualized datasets for 16.9\%, simulation-only datasets for 12.0\%, and hybrid environments for 3.6\%. A further 12.0\% have mixed provenance. This distribution reflects a well-known constraint in ICS security research: operational ICS environments are safety-critical, proprietary, and difficult to instrument during attacks~\cite{conti2021survey,gomez2019generation}.

Nevertheless, the dominance of laboratory-derived evidence introduces a realism gap. Physical and emulated testbeds are valuable because they enable repeatable attack execution, controlled instrumentation, and clearer labeling. However, they may simplify timing behaviour, operator interaction, network heterogeneity, background noise, process dynamics, and maintenance-driven non-stationarity~\cite{ekisa2024virtual,rakas2020review}. Consequently, provenance should not be treated as a secondary descriptor. It directly constrains the operational claims that can be drawn from a reported evaluation result.

This distinction is especially important for IIoT and IoT datasets. Datasets such as TON\_IoT, BoT-IoT, N-BaIoT, Edge-IIoTset, and CICIoT2023 improve scale, attack diversity, and labeling availability, but they do not automatically solve the OT realism problem. They are useful for evaluating cyber-side detection methods and heterogeneous device traffic, yet they cannot replace datasets that expose industrial process semantics and Purdue-layer evidence~\cite{alex2023iotdatasets,elouardi2024hybridcnnllm,Sanjalawe2026}. In this respect, D4 separates dataset availability from operational representativeness.

\subsection{Label Quality and Evaluation Comparability}
\label{S:Comparative:D5}

Dimension \textbf{D5} is comparatively stronger than the other dimensions, but it is also highly fragmented. Multi-class labels appear in more than half of the datasets, and only a small part provides both binary and multi-class labels. Only six datasets are binary-only. Thus, the dominant weakness of \textbf{D5} is not simply label absence, but the uneven meaning, granularity, and comparability of labels across datasets.

The unit of evidence varies substantially. Flow-level labels account for 25~datasets, record-level labels for 24, packet-level labels for 13, and packet/flow labels for only 8, while the remaining datasets use mixed or less common units such as logs, scenarios, images, system calls, or packet/connection records. This fragmentation affects evaluation comparability. A flow-level classifier, a packet-level classifier, and a process-record classifier do not observe the same evidence and should not be interpreted as equivalent even when they report the same accuracy, precision, recall, or F1-score~\cite{koay2023machine,elouardi2024hybridcnnllm}.

Label provenance also varies in practical usefulness. Although 89.2\% of the datasets are coded as live-executed, this only indicates that the attacks were actively performed during data collection. It does not necessarily mean that the released labels identify the precise attack interval, targeted component, affected process variable, onset of physical deviation, or resulting operational consequence. In many datasets, the label merely distinguishes attack from normal observations over a broad capture interval, limiting the ability to relate cyber activity to specific process effects. Prior work therefore emphasizes the need for synchronized event-level annotations, explicit attack logs, target information, and process-consequence labels to support reproducible anomaly interpretation and operational-impact analysis~\cite{gomez2019generation,dobler2025systematic,martins2024comparative,afaq2025publicdatasets}. Thus, \textbf{D5} captures not only whether labels exist, but also whether they provide sufficiently precise and traceable ground truth for credible evaluation.

\subsection{Structural Imbalances and Dataset Gaps}
\label{S:Comparative:Gaps}

The comparative dataset taxonomy exposes three major structural imbalances in the current dataset ecosystem. These should not be interpreted as statistical biases in the narrow ML sense. They are corpus-level coverage biases: systematic skews in what public datasets make observable and, consequently, in what IDS evaluations can credibly test. First, the corpus exhibits \emph{architectural shallowness}. Architecture-aware datasets are concentrated at controller, supervisory, and network-adjacent layers, while field-device and process-interaction evidence remains sparse. This limits the ability of public datasets to support claims about physical-impact detection, because the evidence closest to the physical process is rarely available~\cite{conti2021survey,yadav2021architecture,lin2019using}.

Second, the corpus exhibits \emph{progression compression}. Stage~2 disruption is well represented, but cross-stage intrusion progression is rare. As a result, IDS evaluation is still largely performed on attack episodes rather than realistic campaigns that include intrusion preparation, staging, IT/OT boundary crossing, and process manipulation~\cite{martins2024comparative,dobler2025systematic,koay2023machine}. Third, the corpus exhibits \emph{cross-domain substitution}. IoT and general IDS datasets are abundant, recent, and often well labeled, which makes them attractive for ML evaluation. However, their \textbf{D3}=\textit{N/A} status indicates that they do not encode industrial architectural semantics. They can support general intrusion-detection research, but they cannot close gaps in OT-specific evidence~\cite{alex2023iotdatasets,Sanjalawe2026,raza2026deeplearning}.

Taken together, these imbalances explain why performance results across ICS, IIoT, IoT, and IDS datasets are often difficult to compare. The datasets do not merely differ in size, recency, or number of attack labels; they differ in the kind of evidence they make observable. Therefore, results should not be generalized beyond the evidence profile of the dataset on which they were obtained. Generic intrusion-classification claims may be supported by IoT or IDS baselines, but ICS threat-detection claims require relevant \textbf{D1}, \textbf{D2}, and \textbf{D3} evidence, while operational-impact claims require \textbf{D5} labels that are granular enough to connect cyber events to process consequences. Overall, the taxonomy shows that the dataset ecosystem is broader than it is deep. The field has many datasets, but relatively few combine adversary lifecycle coverage, Purdue-layer evidence, realistic provenance, and high-quality labels. The proposed \textbf{D1--D5} taxonomy in Figure~\ref{fig:unified_taxonomy}, therefore provides a structured basis for interpreting dataset suitability, identifying benchmark gaps, and avoiding overgeneralized claims from narrow evaluation settings.

\section{Evaluation Practices}
\label{S:Evaluation}

The empirical value of a dataset, including ICS-oriented, is ultimately realized through the evaluation protocols applied to it. In other words, coverage, fidelity, and label quality determine what a dataset contains, while evaluation methodology determines what conclusions can legitimately be drawn from it. Across the 18 studies included in this meta-review, the reporting and treatment of evaluation practices are markedly heterogeneous. Specifically, the surveys differ in whether and how they address dataset partitioning, performance metrics, threshold calibration, and related evaluation choices, making it difficult to establish consistent conclusions about evaluation rigor across the literature. Note that these concerns are not novel observations at the level of individual studies; they have been raised repeatedly in ICS dataset and testbed surveys~\cite{hu2018survey,mr2021machine,kamp2023}. What this section contributes is a characterization of those concerns at the meta-survey level, identifying which evaluation weaknesses are pervasive, which are study-specific, and what the combined effect of multiple co-occurring weaknesses implies for the reliability of the reported evidence base.

\begin{table*}[!htbp]
\centering
\caption{Common evaluation design choices observed across the surveyed
  ICS IDS literature, their typical failure modes in ICS operational
  contexts, and recommended practice.}
\label{T:EvalPractices}
\renewcommand{\arraystretch}{1.18}
\begin{tabular}{@{}p{0.17\linewidth}p{0.20\linewidth}p{0.25\linewidth}p{0.3\linewidth}@{}}
\hline
\textbf{Evaluation axis} &
\textbf{Common practice} &
\textbf{Typical failure mode} &
\textbf{Recommended practice} \\
\hline
Data partitioning & Random record-level split; $k$-fold CV & Temporal leakage via autocorrelation and attack adjacency; inflated F1 and AUC & Chronological split with non-overlapping windows; train-only normalization; validation-only tuning \\
\hline
Windowing and overlap & Sliding windows with overlap across partitions & Implicit information sharing across folds; optimistic generalization estimates & Block or grouped splitting; enforce zero window overlap across train, validation, and test \\
\hline
Threshold calibration & Threshold tuned on test data, or tuned per attack scenario & Optimistic operating point; threshold encodes attack knowledge & Calibrate threshold on validation segment reflecting normal variability; lock before final test \\
\hline
Label granularity & Pointwise record-level labels treated as independent samples &  Late detection scores equally with timely detection; early warning penalized; inconsistent semantics & Event-aware scoring for episodic attacks; report both pointwise and event-level outcomes \\
\hline
Metrics reported & Accuracy and F1; occasionally AUC & Accuracy dominated by majority class; detection delay and alarm burden ignored & Report confusion matrices, precision, recall, F1, and AUC; supplement these with the false alarm rate (FAR) per unit time, detection delay, and alarms per attack event \\
\hline
Validation scope & Single dataset, single split & Overfitting to dataset-specific conditions; no evidence of transfer & Multi-dataset evaluation; cross-dataset transfer where feasible; report failure cases explicitly \\
\hline
Evaluation mode & Offline batch evaluation only & Runtime, concept drift, and latency constraints untested & Add streaming or pseudo-online evaluation; report throughput and latency bounds \\
\hline
Reproducibility & Sparse reporting of preprocessing, partition seeds, and calibration details & Results not independently replicable; hidden degrees of freedom & Publish partition scripts, preprocessing pipeline, fixed seeds, threshold calibration procedure, and full configuration \\
\hline
\end{tabular}
\end{table*}

\begin{table*}[!htbp]
\centering
\caption{Evaluation practice coverage across the 18~included studies.
  Scoring: \Full~=~explicitly addressed and well-reported;
  \Partial~=~partially addressed or ambiguously reported;
  \None~=~absent or unreported.
  Column abbreviations correspond to the evaluation axes of
  Table~\ref{T:EvalPractices}.}
\label{T:EvalCoverage}
\renewcommand{\arraystretch}{1.15}
\resizebox{\textwidth}{!}{%
\begin{tabular}{@{}p{0.30\linewidth}C{0.082\linewidth}C{0.082\linewidth}C{0.082\linewidth}C{0.082\linewidth}C{0.082\linewidth}C{0.082\linewidth}C{0.082\linewidth}C{0.082\linewidth}@{}}
\hline
\textbf{Study} &
  \rotatebox{75}{\textbf{Partitioning}} &
  \rotatebox{75}{\textbf{Windowing}} &
  \rotatebox{75}{\textbf{Threshold}} &
  \rotatebox{75}{\textbf{Label granularity}} &
  \rotatebox{75}{\textbf{Metrics}} &
  \rotatebox{75}{\textbf{Val.\ scope}} &
  \rotatebox{75}{\textbf{Eval.\ mode}} &
  \rotatebox{75}{\textbf{Reproducibility}} \\
\hline

Raza et al.~\cite{raza2026deeplearning} (2026)
  & \None & \None & \Partial & \None & \Full & \None & \None & \None \\

Sanjalawe et al.~\cite{Sanjalawe2026} (2026)
  & \Partial & \None & \None & \None & \Full & \None & \None & \None \\

Afaq et al.~\cite{afaq2025publicdatasets} (2025)
  & \Full & \Partial & \Partial & \Partial & \Full & \Full & \None & \Partial \\

Dobler et al.~\cite{dobler2025systematic} (2025)
  & \Full & \Full & \Full & \Full & \Full & \Partial & \None & \Full \\

Muhammad et al.~\cite{muhammad2024smartgrid} (2024)
  & \Partial & \None & \None & \None & \Full & \None & \None & \None \\

Elouardi et al.~\cite{elouardi2024hybridcnnllm} (2024)
  & \Partial & \Partial & \Partial & \None & \Full & \None & \None & \None \\

Martins et al.~\cite{martins2024comparative} (2024)
  & \Full & \Partial & \Partial & \Partial & \Full & \Full & \None & \Partial \\

Ekisa et al.~\cite{ekisa2024virtual} (2024)
  & \None & \None & \None & \None & \Partial & \None & \None & \None \\

Alanazi et al.~\cite{alanazi2023scada} (2023)
  & \Partial & \None & \None & \None & \Full & \None & \None & \None \\

Alex et al.~\cite{alex2023iotdatasets} (2023)
  & \Partial & \None & \None & \None & \Full & \Partial & \None & \None \\

Alani et al.~\cite{alani2023survey} (2023)
  & \Partial & \None & \None & \None & \Full & \None & \None & \None \\

Koay et al.~\cite{koay2023machine} (2023)
  & \Full & \Full & \Partial & \Partial & \Full & \Partial & \None & \Partial \\

Yadav \& Paul~\cite{yadav2021architecture} (2021)
  & \None & \None & \None & \None & \Partial & \None & \None & \None \\

Mubarak et al.~\cite{mubarak2021anomaly} (2021)
  & \Partial & \Partial & \Partial & \Full & \Full & \None & \None & \Partial \\

Conti et al.~\cite{conti2021survey} (2021)
  & \Full & \Partial & \Partial & \Partial & \Full & \Partial & \None & \Partial \\

Rakas et al.~\cite{rakas2020review} (2020)
  & \Partial & \None & \None & \None & \Partial & \None & \None & \None \\

Lin et al.~\cite{lin2019using} (2019)
  & \Full & \Full & \Full & \Full & \Full & \Full & \None & \Full \\

G\'omez et al.~\cite{gomez2019generation} (2019)
  & \Full & \Partial & \Partial & \Full & \Full & \None & \None & \Partial \\

\hline
\multicolumn{9}{@{}p{\dimexpr0.956\linewidth+16\tabcolsep}@{}}{%
\footnotesize\textbf{Column totals:}
Partitioning~(\Full:~7, \Partial:~8, \None:~3);\;
Windowing~(\Full:~3, \Partial:~6, \None:~9);\;
Threshold~(\Full:~2, \Partial:~8, \None:~8);\;
Label granularity~(\Full:~4, \Partial:~4, \None:~10);\;
Metrics~(\Full:~15, \Partial:~3, \None:~0);\;
Val.\ scope~(\Full:~3, \Partial:~4, \None:~11);\;
Eval.\ mode~(\Full:~0, \Partial:~0, \None:~18);\;
Reproducibility~(\Full:~2, \Partial:~6, \None:~10).} \\
\hline
\end{tabular}}
\end{table*}

\begin{figure*}[!ht]
  \centering
  \includegraphics[width=\linewidth]{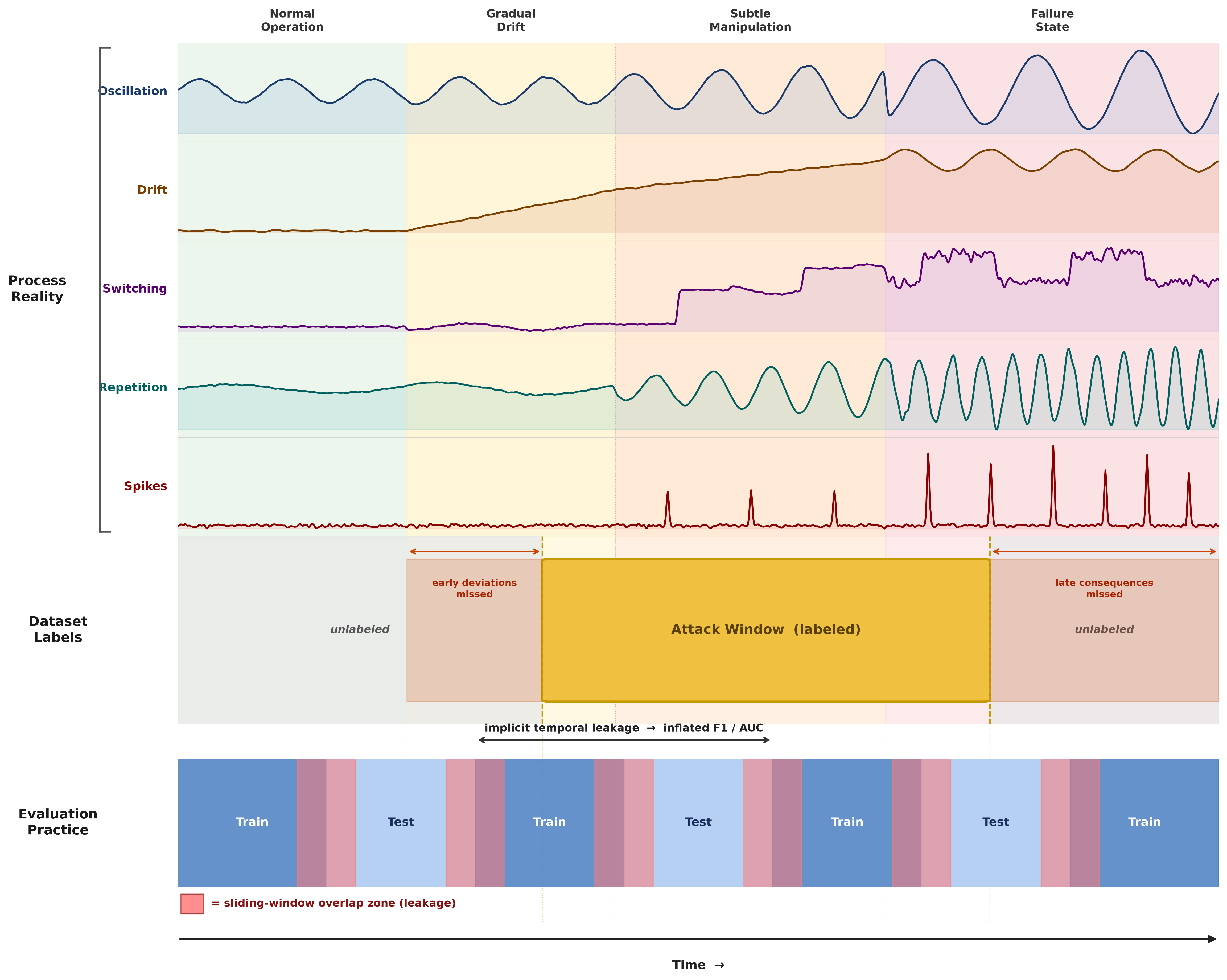}
  \caption{Three structural sources of evaluation bias in ICS IDS research are exposed by aligning process reality, dataset labeling, and temporal splitting on a common time axis.}
  \label{F:EvalTimeline}
\end{figure*}

Figure~\ref{F:EvalTimeline} visualizes the typical three structural sources of evaluation bias in ICS IDS to facilitate the subsequent analysis of sections~\ref{SS:eval_partitioning} to~\ref{SS:eval_reproducibility}. Its strata, aligned on a common time axis, capture a structural misalignment that recurs systematically across the surveyed literature. The top group of five rows represents the physical reality of process evolution: each row depicts one characteristic ICS anomaly pattern across the four operational phases; recall that industrial processes can be manipulated in many ways, including gradual sensor drifts, abrupt spikes, frozen telemetry, oscillatory control behavior, and forced actuator switching~\cite{kampourakis2026}. Driven by physical inertia, control feedback, and operator interaction, these phenomena may co-occur and evolve continuously, rarely conforming to the abrupt, discrete event boundaries that pointwise labeling schemes implicitly assume.

The middle stratum depicts the corresponding dataset labeling practice. Namely, a single attack window is assigned that only partially overlaps with the underlying process degradation. For example, as depicted in figure~\ref{F:EvalTimeline}, the labeled onset falls after drift has already begun and after the first switching events have occurred, leaving those phenomena classified as normal; the labeled offset precedes the end of physical consequences, so post-spike instability is either unlabeled or misclassified. This onset/offset mismatch, made spatially explicit by the red mismatch spans in the figure, may obscure the boundary between early warning and late detection, and the phenomena that may be the most informative for timely intervention are those excluded from the labeled region.

Finally, the bottom stratum conceptually illustrates how dataset labels may interact with evaluation design. Random or weakly constrained temporal splits, particularly when combined with overlapping windows or sliding feature extraction, can allow temporally adjacent samples from the same process scenario to appear in both training and testing partitions. In Figure~\ref{F:EvalTimeline}, the pink overlap zones illustrate this potential source of implicit leakage. Such leakage can inflate pointwise metrics such as F1-score and AUC, reward late or repeated detections, and obscure operationally relevant failure modes, including delayed response to drift, alarm bursts triggered by repetition, and sensitivity to spike transients. The five subsections below examine these evaluation issues in more detail, using the dimensions summarized in Table~\ref{T:EvalPractices} and the survey-level coverage reported in Table~\ref{T:EvalCoverage}.

\subsection{Data Partitioning and Temporal Leakage}
\label{SS:eval_partitioning}

A recurring weakness across the surveyed works is the use of random record-level splits and standard $k$-fold cross-validation on temporally correlated ICS traces~\cite{raza2026deeplearning,ekisa2024virtual,yadav2021architecture,alanazi2023scada}. In process datasets and controlled testbed traffic, adjacent sensor readings, actuator states, and network payloads are strongly autocorrelated, while attack episodes occur in contiguous intervals rather than as independently distributed events. Random shuffling can therefore place near-duplicate temporal context in both training and test sets. Experimental work shows that such leakage can produce F1-score and AUC estimates substantially higher than those obtained under chronological partitioning, with the gap increasing with autocorrelation and feature-window size~\cite{sasse2025overview}. Consequently, strong pointwise performance under leakage may not translate to reliable detection on continuous operational traces.

Chronological partitioning is more appropriate for ICS time series because it preserves control-cycle structure, gradual drift, and realistic scenario progression. However, temporal splitting alone does not guarantee leakage-free evaluation: sliding windows may overlap partition boundaries, normalization may be fitted globally, thresholds may be calibrated on test data, or hyperparameters may be tuned without a temporally separated validation set~\cite{conti2021survey,martins2024comparative,afaq2025publicdatasets}. These choices can independently introduce optimistic bias, while their incomplete reporting makes methodological quality difficult to assess~\cite{khraisat2019survey}. Table~\ref{T:EvalCoverage} records which partitioning details are reported by each included study.

\subsection{Ground-Truth Representation and Label Granularity}
\label{SS:eval_labels}

The predominant evaluation paradigm treats ICS intrusion detection as a pointwise classification problem, where each record or window is independently labeled as normal or attack. This framing can be misaligned with the operational reality of ICS attacks because labeled attack windows do not necessarily coincide with the underlying physical-process degradation~\cite{raza2026deeplearning,alanazi2023scada,alani2023survey,muhammad2024smartgrid}. As illustrated in Figure~\ref{F:EvalTimeline}, early deviations may precede the nominal label onset, while physical consequences may continue beyond the labeled offset. Pointwise evaluation can therefore penalize early warning or fail to distinguish detection of the process event from detection of the predefined label boundary~\cite{chaabouni2019network}.

ICS attacks are also scenario-driven and episodic, spanning contiguous intervals in which detection timing matters. Under pointwise scoring, a detector that first raises an alarm substantially after attack onset can achieve the same recall as one that detects the event early, provided both cover a similar portion of the labeled interval. This latency-blindness is problematic where detection delay has direct safety and reliability consequences~\cite{shin2021two,morris2011control}. Event-based ground truth with documented onset and offset timestamps enables event-aware scoring, timely first-detection assessment, repeated-alarm penalties, and detection-delay measurement. Nevertheless, event-aware evaluation remains rare, and studies frequently reduce datasets with attack timestamps to record-level accuracy and F1~\cite{alanazi2023scada,alani2023survey,alex2023iotdatasets,rakas2020review,conti2021survey}.

A further limitation is that heterogeneous attack behaviours are often collapsed into a common binary ``attack'' label. As illustrated in Figure~\ref{F:EvalTimeline}, attacks may exhibit different temporal morphologies, including spikes, drifts, oscillations, repetition, and switching patterns~\cite{kampourakis2026}. Binary ground truth can therefore obscure both timing information and intra-class behavioral variation. This problem is directly related to the \textbf{D5} dimension of the unified taxonomy: datasets with coarse or run-level labels constrain event-aware evaluation regardless of the detection method applied~\cite{yadav2021architecture,ekisa2024virtual,raza2026deeplearning,Sanjalawe2026}. Weak label granularity consequently limits the ability to evaluate timely detection, behavioral robustness, and operationally meaningful attack outcomes.

\subsection{Metric Selection and Operational Alignment}
\label{SS:eval_metrics}

Metric selection across the surveyed corpus is inconsistent and often poorly aligned with the operational context of ICS intrusion detection. Accuracy remains reported despite substantial class imbalance, where high scores may provide little security value~\cite{alshamrani2019survey,yadav2021architecture,rakas2020review,ekisa2024virtual}. Precision, recall, F1-score, and ROC AUC are more informative and more commonly reported~\cite{lin2019using,dobler2025systematic,koay2023machine,mubarak2021anomaly}, but remain time- and cost-agnostic. Detectors with similar F1-scores may differ substantially in detection delay and alarm burden~\cite{caselli2015sequence}; therefore, ICS-focused evaluation should also consider detection delay, false alarm rate per unit time, and alert burden per event~\cite{alshamrani2019survey,kampourakis2025balancing}. However, such measures remain rarely and inconsistently reported across the surveyed corpus~\cite{raza2026deeplearning,muhammad2024smartgrid,alanazi2023scada,ekisa2024virtual}.

A related issue concerns threshold calibration. Continuous-output detectors require a threshold to produce alerts, and precision, recall, and F1 are sensitive to this choice. When thresholds are selected post-hoc on test data rather than calibrated on a held-out validation segment, the reported operating point may reflect evaluation design rather than detector quality~\cite{pinto2023survey,elouardi2024hybridcnnllm,conti2021survey,martins2024comparative}. Thus, threshold leakage and the partitioning weaknesses of Section~\ref{SS:eval_partitioning} are closely related.

\subsection{Generalization Evidence and Transfer Validity}
\label{SS:eval_transfer}

Most evaluations in the surveyed corpus rely on a single dataset and a single split~\cite{raza2026deeplearning,alanazi2023scada,alani2023survey,ekisa2024virtual,yadav2021architecture}. This provides limited evidence that reported performance generalizes beyond the specific evaluation setting. This is particularly important because ICS datasets differ in process type, protocol stack, attack distribution, label granularity, and testbed fidelity. Cross-dataset or multi-benchmark evaluation provides stronger evidence of generalization, but remains uncommon~\cite{afaq2025publicdatasets,martins2024comparative,lin2019using}. Where transfer is examined, performance typically degrades relative to within-dataset results, consistent with the \textbf{D3} and \textbf{D4} differences identified by the unified taxonomy~\cite{mr2021machine}.

Generalization is further limited by the prevalence of offline batch evaluation. Although suitable for controlled comparison, offline evaluation does not assess concept drift, threshold stability under operational variability, or computational latency under real-time throughput constraints~\cite{mr2021machine,shyaa2023enhanced}. None of the surveyed works report streaming or pseudo-online evaluation that approximates deployed ICS operation~\cite{dobler2025systematic,lin2019using,koay2023machine}.

\subsection{Reproducibility and Reporting Completeness}
\label{SS:eval_reproducibility}

Reproducible ICS intrusion-detection evaluation requires sufficient reporting of dataset partitions, preprocessing and feature-extraction procedures, threshold calibration, hyperparameter tuning, validation data, and dataset versions or cleaning steps. Across the surveyed corpus, reporting on these dimensions remains sparse and heterogeneous~\cite{raza2026deeplearning,alanazi2023scada,alani2023survey,ekisa2024virtual,yadav2021architecture,rakas2020review}. Studies using established benchmarks generally provide more procedural detail~\cite{lin2019using,dobler2025systematic}, whereas evaluations involving less-established or repurposed datasets more frequently omit information needed for independent replication~\cite{rakas2020review,ekisa2024virtual,yadav2021architecture,pinto2023survey}.

This weakness is also connected to \textbf{D4}: poorly documented or weakly maintained datasets propagate methodological ambiguity into downstream evaluations. Therefore, reproducibility  depends on both rigorous study-level protocols and better-documented, better-maintained datasets at the ecosystem level.

\subsection{Taxonomy--Evaluation Coupling}
\label{SS:eval_taxonomy_coupling}

Taken together, Sections~\ref{SS:eval_partitioning} to~\ref{SS:eval_reproducibility} show that evaluation weaknesses are not independent of the structural limitations captured by the unified taxonomy. As Tables~\ref{T:EvalPractices} and~\ref{T:EvalCoverage} indicate, random splitting, pointwise labels, conventional classification metrics, single-dataset validation, and offline evaluation frequently occur together. Improvements along one evaluation axis may therefore remain constrained by weaknesses elsewhere, particularly the interaction between label granularity (\textbf{D5}), partitioning discipline, and metric selection.

\smallskip\noindent\textbf{D1 (Adversarial Tactic Coverage) and D2 (Attack Progression Depth) constrain evaluation validity.}
Datasets covering only a narrow portion of the ATT\&CK for ICS tactic space support performance claims only within that subspace. Stage~2-only datasets provide no evidence of detection capability for Stage~1 activities such as reconnaissance, delivery, or lateral movement, while the scarcity of realistic cross-stage datasets constrains multi-stage evaluation~\cite{dobler2025systematic,afaq2025publicdatasets,koay2023machine}.

\smallskip\noindent\textbf{D3 (System Scope and Evidence Layer) affects transfer validity.}
Datasets concentrated at Purdue Levels~2--3 cannot directly support evaluation of methods requiring Level~0--1 field-device evidence. Cross-dataset transfer across different architectural layers is therefore limited not only by covariate shift but also by differences in the observable evidence itself, helping explain the transfer gap discussed in Section~\ref{SS:eval_transfer}.

\smallskip\noindent\textbf{D4 (Operational Realism and Provenance) constrains deployment-oriented evaluation.}
Simulation, emulation, and controlled testbeds provide limited evidence about properties such as streaming latency, concept drift, and operational variability. Consequently, even improved evaluation protocols cannot fully characterize deployment performance when operationally sourced data remain scarce.

\smallskip\noindent\textbf{D5 (Detection Evidence Quality) constrains event-aware evaluation.}
Coarse or run-level labels restrict the measurement of detection delay, alarm burden, and early warning when precise attack boundaries are unavailable. Thus, the applicability of event-aware and delay-aware metrics depends directly on the temporal and semantic precision of the available ground truth.

Overall, the taxonomy--evaluation coupling shows that methodological improvement cannot be separated from dataset improvement. More rigorous partitioning, metric selection, and validation can strengthen evaluation, but their credibility remains bounded by the adversarial coverage, architectural scope, provenance, and ground-truth quality of the datasets on which they are applied.

\begin{table*}[!t]
\centering
\caption{Coupling between unified taxonomy dimensions (D1--D5) and evaluation practices.}
\label{T:TaxoCouplingMap}
\renewcommand{\arraystretch}{1.18}
\begin{tabular}{@{}p{0.07\linewidth}p{0.26\linewidth}p{0.32\linewidth}p{0.22\linewidth}@{}}
\hline
\textbf{Dim.} & \textbf{Dataset-level limitation} & \textbf{Evaluation consequence} & \textbf{Affected eval.\ axes} \\
\hline
D1 & Tactic coverage concentrated in OT Disruption and Impact; Recon.\ and Staging largely isolated from continuous attack progression & Evaluation valid only within a narrow tactic subspace; no evidence of Stage~1 detection capability & Validation scope; generalization \\
\hline
D2 & Cross-stage IT/OT datasets are rare; Stage 2-only datasets dominate & Cross-stage detector evaluation is severely constrained by the limited evidence base; ceiling on benchmark ambition & Validation scope; generalization \\
\hline
D3 & Evidence concentrated at Purdue Levels~2--3; Level~0--1 sparse & Cross-dataset transfer gap is partly architectural, not reducible by better partitioning & Partitioning; generalization \\
\hline
D4 & Operational datasets are rare; most datasets originate from controlled physical, simulated, or emulated environments & Limited operational data constrains streaming, latency, and concept-drift evaluation & Evaluation mode; reproducibility \\
\hline
D5 & Fine-grained event-level onset/offset annotations are rare; many datasets rely on coarser label schemes & Event-aware evaluation, delay measurement, and alarm-burden metrics are constrained when event boundaries are unavailable & Label granularity; metrics; reproducibility \\
\hline
\end{tabular}
\end{table*}

Table~\ref{T:TaxoCouplingMap} makes the evaluation practices and unified taxonomy of section~\ref{S:unified_taxonomy} coupling explicit by mapping each of the taxonomy's dimensions to the evaluation axes it constrains, identifying the direction of the dependency, and noting which studies are jointly affected by both the taxonomy- and evaluation-level weakness. The key takeaway is that addressing evaluation methodology in isolation, improving partitioning or metric selection without simultaneously improving the datasets on which evaluation is performed, can yield incomplete and potentially misleading improvements. Conversely, releasing better-labeled, more architecturally diverse datasets without updating community evaluation norms can not be translated into more credible reported results. Essentially, progress requires coordinated improvement at both levels, with taxonomy-evaluation coupling providing a solid analytical basis for prioritizing which improvements can have the greatest downstream effect.

\section{The Road Ahead}
\label{S:RoadAhead}

The corpus characterised in this study, namely, 83 datasets drawn (section~\ref{S:comparative_dataset_analysis}) from 18 studies (section~\ref{S:overview}), provides a comprehensive empirical baseline for assessing the state of ICS cybersecurity datasets.  The analysis reveals not merely isolated gaps but three structurally interlocking deficits that constrain what the research community can claim about ICS intrusion detection: i) an architectural shallowness that leaves field-device and process-interaction layers almost entirely unevidenced, ii) a progression compression that privileges late-stage OT disruption while cross-stage IT/OT intrusion campaigns remain essentially invisible, and iii) a cross-domain substitution that inflates apparent dataset abundance, while contributing no OT-specific signal. These imbalances, quantified in Section~\ref{S:comparative_dataset_analysis} and linked to their evaluation consequences in Table~\ref{T:TaxoCouplingMap}, cannot be resolved by refining evaluation methodology alone. Coordinated progress on all three fronts, i.e., dataset construction, evaluation design, and shared community infrastructure, is necessary. The discussion below articulates that agenda with specificity proportional to the evidence gathered here.

\subsection{Priorities for Dataset Creators}
\label{S:RoadAhead:Datasets}

The taxonomy dimensions \textbf{D1--D5} (Section~\ref{S:unified_taxonomy}) reveal a coherent pattern of under-investment. With 85.5\% of the 83~datasets concentrated in OT~Disruption tactics (\textbf{D1}) and 57.8\% limited to Stage~2 progression depth (\textbf{D2}), the corpus primarily captures the final phases of intrusion campaigns. Although Reconnaissance and Staging appears in 55.4\% of datasets, it is usually represented as isolated telemetry rather than as part of a continuous sequence crossing the IT/OT boundary. Future dataset construction should therefore treat cross-stage coverage as a primary design objective.

Architectural and provenance limitations are equally important. Under \textbf{D3}, 50.6\% of datasets are not applicable to ICS system-scope semantics, while Level~0 field-device evidence is effectively absent and Levels~2--3 dominate OT-specific coverage (Figure~\ref{fig:taxonomy_distribution}). Future collection should therefore target lower Purdue levels~\cite{iec62443}, including serial-bus traffic, physics-layer process measurements, and firmware-level instrumentation. Under \textbf{D4}, physical testbeds account for 39.8\% of datasets, while only 15.7\% are operationally sourced and simulation/emulation together account for 28.9\%. Since transfer from simulated or emulated environments to physical or operational settings can degrade performance~\cite{conti2021survey,dobler2025systematic}, promising directions include federated data-sharing agreements and hybrid provenance designs that inject synthetic attacks into operational background traffic~\cite{gomez2019generation,ekisa2024virtual}.

Finally, \textbf{D5} remains central to evaluation comparability. Although live-executed labels dominate (89.2\%), label granularity is often coarse and event-level onset and offset annotations remain uncommon, while multi-class schemes, present in 54.2\% of datasets, are rarely aligned across studies. A community-agreed label schema, analogous to MITRE ATT\&CK for ICS~\cite{mitre_ics} but extended with temporal event boundaries and sensor-level attribution, could improve cross-dataset comparability and provide a more consistent basis for reproducible evaluation.

\subsection{Priorities for Evaluation Design}
\label{S:RoadAhead:Evaluation}

The evaluation audit in Section~\ref{S:Evaluation} reveals substantial methodological limitations. None of the 18~included studies report streaming or online evaluation, partitioning discipline is adequate in only 7, and reproducibility is fully reported by only two (Table~\ref{T:EvalCoverage}). The coupling analysis in Table~\ref{T:TaxoCouplingMap} links these weaknesses to dataset limitations. Scarce cross-stage datasets \textbf{(D2)} constrain cross-stage evaluation, limited operational data \textbf{(D4)} restrict streaming, latency, and concept-drift assessment, and coarse labels \textbf{(D5)} limit event-aware and delay-sensitive metrics. Three priorities follow.

\begin{itemize}

    \item \textbf{Transfer-oriented benchmarking.} Given the architectural heterogeneity of ICS environments, strong performance on a single dataset provides limited assurance. Evaluation across datasets with different provenance classes (\textbf{D4}), system-scope layers (\textbf{D3}), or protocol contexts provides stronger evidence of generalization. Transfer evaluation should therefore become a reporting expectation rather than an optional extension.

    \item \textbf{Temporal realism and streaming protocols.} The absence of streaming evaluation reflects both limited availability of online-ready datasets and the lack of common protocols for measuring latency, concept drift, and alarm burden. Addressing this requires event-level annotations, timestamp-preserving formats, realistic inter-arrival distributions, non-stationary train/test splits, and streaming-oriented evaluation conventions. As Figure~\ref{F:EvalTimeline} illustrates, temporal protocols are needed to expose early deviation misses and late consequence detection that static evaluation can conceal.

    \item \textbf{Reproducibility as a minimum reporting standard.} With only 2 of 18 studies fully reporting reproducibility, consistent comparison remains difficult. A reproducibility checklist covering dataset accessibility, preprocessing code, model hyperparameters, random seeds, and partitioning procedures should therefore become a minimum requirement. Shared evaluation harnesses using canonical preprocessing and fixed dataset splits could further reduce implementation-driven variability.
    
\end{itemize}

\subsection{Shared Infrastructure and Governance}
\label{S:RoadAhead:Infrastructure}

Individual dataset and methodological improvements remain fragmented without shared infrastructure that enables cumulative progress. Three mechanisms are particularly important.

\begin{itemize}

    \item \textbf{Federated dataset registry.} A community-maintained ICS cybersecurity registry, analogous to the UCI Machine Learning Repository~\cite{UCI}, could assign persistent dataset identifiers, track versions and corrections, record provenance and licence information, and expose the \textbf{D1--D5} metadata required for dataset selection. Such a registry would also make corpus-level gaps continuously visible and help direct future dataset development toward underrepresented areas.

    \item \textbf{Community benchmark suites.} Stable benchmark suites have supported systematic comparison in other domains, including KITTI for autonomous driving~\cite{KITTI} and GLUE/SuperGLUE for NLP~\cite{NEURIPS2019_4496bf24}. An analogous ICS benchmark could provide reproducible dataset splits spanning multiple system-scope layers (\textbf{D3}) and provenance classes (\textbf{D4}), together with shared metrics that include detection delay and alarm burden alongside conventional classification measures. The \textbf{D1--D5} taxonomy provides a natural structure for organizing such benchmark coverage.

    \item \textbf{Ethical and operational constraints.} Increasing the use of operational ICS data creates privacy, commercial, and safety concerns because traces may expose process parameters, production schedules, or safety-system configurations. Explicit data-sharing frameworks should therefore define acceptable anonymisation procedures, retention requirements, and conditions under which synthetic augmentation can substitute for operational disclosure. Without such governance, the \textbf{D4} gap between simulation-derived and operational datasets is likely to persist.
    
\end{itemize}

\subsection{Research Agenda}
\label{S:RoadAhead:Agenda}

Building on the structural gaps identified above, three research directions appear particularly important.

\begin{itemize}

    \item \textbf{Cross-stage dataset construction methodology.} Existing datasets rarely capture a complete IT/OT intrusion sequence from reconnaissance to process disruption because Stage~1 activity and Stage~2 physical effects occur across architecturally distinct environments. Research should therefore examine digital-twin and cyber-range architectures~\cite{KampourakisDT2025,kamp2025}, co-simulation frameworks, and instrumented red-team exercises capable of bridging this boundary and supporting end-to-end dataset capture~\cite{ekisa2024virtual,gomez2019generation,lin2019using}.

    \item \textbf{Evaluation protocols for non-stationary ICS telemetry.} ICS signals exhibit seasonality, maintenance-driven non-stationarity, and gradual degradation that standard IID assumptions~\cite{IID} do not capture. Temporally structured evaluation, expanding-window validation, concept-drift benchmarks, and realistic alarm-rate constraints are needed to connect reported detection performance with deployment conditions. As illustrated in Figure~\ref{F:EvalTimeline}, different temporal phenomena such as oscillation, drift, switching, repetition, and spikes can otherwise be conflated under static evaluation.

    \item \textbf{Metadata standards for taxonomy-aligned characterisation.} The \textbf{D1--D5} taxonomy provides a common characterization structure, but its usefulness depends on consistent annotation. Structured reporting templates, NLP-assisted harmonisation, and LLM-aided extraction from dataset documentation could reduce the cost of maintaining taxonomy metadata and support the registry proposed in Section~\ref{S:RoadAhead:Infrastructure}. The alignment of such metadata with MITRE ATT\&CK for ICS~\cite{mitre_ics}, IEC~62443~\cite{iec62443}, and NIST SP~800-82r3~\cite{nist80082} would avoid introducing unnecessary new terminology and preserve compatibility with established frameworks.
    
\end{itemize}

\section{Conclusion}
\label{S:conclusion}

This paper set out to examine whether the collective body of public experimental data supports the evaluation claims made in ICS cybersecurity research. Based on the systematic characterisation of 83 datasets identified across 18 studies, the answer is negative in several measurable respects. The five-dimensional taxonomy developed in Section~\ref{S:unified_taxonomy} shows that the corpus is structurally skewed: 85.5\% of datasets concentrate on OT~Disruption (\textbf{D1}), 57.8\% capture Stage~2 activity in isolation and only 8.4\% span multiple stages (\textbf{D2}), Level~0 field-device evidence is effectively absent (\textbf{D3}), and operationally sourced data accounts for only 15.7\% of the corpus (\textbf{D4}). Although 89.2\% of datasets carry live-executed labels, event-level onset and offset annotations remain rare (\textbf{D5}).

These weaknesses also constrain evaluation practice. The taxonomy-evaluation coupling analysis in Table~\ref{T:TaxoCouplingMap} shows that scarce cross-stage coverage limits cross-stage evaluation, limited operational data restricts streaming and concept-drift assessment, and coarse ground truth constrains delay-aware and alarm-burden evaluation. The audit in Section~\ref{S:Evaluation} reinforces this finding: none of the 18 included studies report streaming evaluation, only two satisfy reproducibility requirements, and partitioning discipline is adequate in fewer than half. Consequently, benchmark performance claims are often bounded by the structural properties of the datasets on which they are obtained.

The contribution of this meta-review is therefore threefold. First, it provides a unified characterisation of a fragmented dataset corpus using a standard-anchored \textbf{D1--D5} taxonomy. Second, it establishes a quantitative baseline through the distributions in Figure~\ref{fig:taxonomy_distribution}, the evaluation audit in Table~\ref{T:EvalCoverage}, and the coupling map in Table~\ref{T:TaxoCouplingMap}. Finally, it translates these findings into a structured research agenda that covers cross-stage dataset construction, temporally structured evaluation, event-level label standards, shared benchmark infrastructure, and a community-maintained registry indexed by \textbf{D1--D5} metadata.

The fundamental argument is that dataset adequacy and evaluation validity must be assessed jointly. A well-labelled but architecturally shallow dataset cannot support claims about field-device detection, while rigorous evaluation on Stage~2-only data cannot establish capability against early-stage intrusion. In other words, a narrow corpus produces narrow evaluation, and narrow evaluation can conceal the narrowness of the corpus. Addressing this cycle requires coordinated progress in dataset construction, evaluation methodology, shared benchmarking, and governance for operational data sharing, providing a stronger empirical foundation for future ICS cybersecurity research.

\subsection*{Acknowledgments}
This work is supported by the Research Council of Norway through
the SFI Norwegian Centre for Cybersecurity in Critical Sectors (NORCICS) project no. 310105

\appendix
\section{Search Queries Used in Each Database}
\label{app:queries}

This appendix lists the database-specific search strings used to retrieve the initial corpus of records for the meta-review. The searches were last executed on June 18, 2026. The queries were adapted to the syntax of each platform while preserving the same conceptual structure across databases. In all cases, the search combined four blocks: (i) domain terms related to industrial control and operational technology environments, (ii) dataset- and benchmark-related terms, (iii) cybersecurity-related terms, and (iv) review-oriented publication terms.

\subsection{Web of Science}
\begin{quote}\small
\texttt{AB=(("industrial control system" OR ICS OR SCADA OR OT OR IIoT OR
"operational technology" OR "critical infrastructure") AND
(dataset OR benchmark OR testbed OR "data collection") AND
("cybersecurity" OR "intrusion detection" OR "anomaly detection" OR
attack OR IDS) AND (survey OR review OR "meta-analysis" OR
"systematic review" OR "literature review"))}
\end{quote}

\subsection{IEEE Xplore}
\begin{quote}\small
\texttt{("Abstract":"Industrial Control System" OR "Abstract":"ICS" OR
"Abstract":"SCADA" OR "Abstract":"OT" OR "Abstract":"IIoT" OR
"Abstract":"Operational Technology" OR "Abstract":"Critical Infrastructure") AND
("Abstract":"dataset" OR "Abstract":"benchmark" OR
"Abstract":"testbed" OR "Abstract":"data collection") AND
("Abstract":"cybersecurity" OR "Abstract":"intrusion detection" OR
"Abstract":"anomaly detection" OR "Abstract":"attack" OR
"Abstract":"IDS") AND ("Abstract":"survey" OR "Abstract":"review" OR
"Abstract":"meta-analysis" OR "Abstract":"systematic review" OR
"Abstract":"literature review")}
\end{quote}

\subsection{ACM Digital Library}
\begin{quote}\small
\texttt{[[Abstract: "industrial control system"] OR [Abstract: "ics"] OR
[Abstract: "scada"] OR [Abstract: "ot"] OR [Abstract: "iiot"] OR
[Abstract: "operational technology"] OR [Abstract: "critical infrastructure"]]
AND [[Abstract: "dataset"] OR [Abstract: "benchmark"] OR
[Abstract: "testbed"] OR [Abstract: "data collection"]]
AND [[Abstract: "cybersecurity"] OR [Abstract: "intrusion detection"] OR
[Abstract: "anomaly detection"] OR [Abstract: "attack"] OR [Abstract: "ids"]]
AND [[Abstract: "survey"] OR [Abstract: "review"] OR
[Abstract: "meta-analysis"] OR [Abstract: "systematic review"] OR
[Abstract: "literature review"]]}
\end{quote}

\subsection{Scopus}
\begin{quote}\small
\texttt{TITLE-ABS-KEY("industrial control system" OR "ics" OR "scada" OR
"ot" OR "iiot" OR "operational technology" OR
"critical infrastructure") AND TITLE-ABS-KEY("dataset" OR "benchmark" OR
"testbed" OR "data collection") AND TITLE-ABS-KEY("cybersecurity" OR
"intrusion detection" OR "anomaly detection" OR "attack" OR
"ids") AND TITLE-ABS-KEY("survey" OR "review" OR "meta-analysis" OR
"systematic review" OR "literature review")}
\end{quote}

\printbibliography

\end{document}